\documentclass[twocolumn]{aastex62}

\usepackage{natbib}
\usepackage{amsmath}
\usepackage{xcolor}

\hypersetup{linkcolor=blue,citecolor=blue,filecolor=blue,urlcolor=blue}

\graphicspath{{./}{figures/}}

\submitjournal{ApJ}

\shorttitle{BLR in 3-D FRADO}
\shortauthors{Naddaf et al.}

\begin{document}

\title{The picture of BLR in 2.5-D FRADO: Dynamics \& Geometry}

\correspondingauthor{Mohammad-Hassan Naddaf}
\email{naddaf@cft.edu.pl}

\author[0000-0002-7604-9594]{Mohammad-Hassan Naddaf}
\affil{Center for Theoretical Physics, Polish Academy of Sciences, Lotnikow 32/46, 02-668 Warsaw, Poland}
\affil{Nicolaus Copernicus Astronomical Center, Polish Academy of Sciences, Bartycka 18, 00-716 Warsaw, Poland}

\author[0000-0001-5848-4333]{Bo\.zena Czerny}
\affil{Center for Theoretical Physics, Polish Academy of Sciences, Lotnikow 32/46, 02-668 Warsaw, Poland}

\author[0000-0003-2886-4460]{Ryszard Szczerba}
\affil{Nicolaus Copernicus Astronomical Center, Polish Academy of Sciences, Rabianska 8, 87-100 Torun, Poland}

\begin{abstract}

The dynamics of the Broad Line Region (BLR) in Active galaxies is an open question, direct observational constraints suggest a predominantly Keplerian motion, with possible traces of inflow or outflow. In this paper we study in detail the physically motivated BLR model of \citet{Czerny2011} based on the radiation pressure acting on dust at the surface layers of accretion disk (AD). We consider here a non-hydrodynamical approach to the
dynamics of the dusty cloud under the influence of radiation coming from the entire AD. We use here the realistic description of the dust opacity, and we introduce two simple geometrical models of the local shielding of the dusty cloud. We show that the radiation pressure acting on dusty clouds is strong enough to lead to dynamical outflow from the AD surface, so the BLR has a dynamical character of (mostly failed) outflow. The dynamics strongly depend on the Eddington ratio of the source. Large Eddington ratio sources show a complex velocity field and large vertical velocities with respect to the AD surface, while for lower Eddington ratio sources vertical velocities are small and most of the emission originates close to the AD surface. Cloud dynamics thus determines the 3-D geometry of the BLR.
\end{abstract}

\keywords{Active Galaxies, Accretion Disk, Radiation Pressure, Broad Line Region, FRADO Model, Shielding Effect, Dust Opacity}

\section{Introduction} \label{sec:intro}

Broad emission lines are the most characteristic features of the optical/UV spectra of most active galaxies. They were the motivation of the study by \citet{seyfert1943}, and they provided the clue to the puzzle of quasar phenomenon \citep{schmidt1963}. The intense emission of BLR from X-ray to optical to infra-red has been studied for years \citep{Boroson1992, Lawrence1997, Sulentic2000, Reeves2000, Gaskell2009, Le2019, Raimundo2020}. These studies rely on the analysis of the spectra, and their time dependence since the region is messy, turbulent and not yet easily spatially resolvable. Line shapes and the photoionization conditions of their formation imply that BLR emission consist of two line categories: High Ionization Lines (HIL) and Low Ionization Lines (LIL) \citep{CollinSouffrin1988}.
Further research brought wealth of information on the spatial distribution of the medium, its motion, and the thermal and ionization state of the material \citep[for a detailed review, see][]{netzer2013} but did not bring the definite answer to the question of the origin of the material located in the BLR.

Reverberation mapping (RM) approach based on the delayed response of BLR material to variation of AGN continuum \citep{Blandford1982, Peterson1993}, enabled us to precisely measure the spectral-dependent radius of BLR \citep{Wandel1999}. It also brought us an insight into the motion of BLR \citep{Peterson1998}, showing that its dynamics is a composition of a Keplerian motion around the BH, and a turbulent motion which overally gives a complex velocity field \citep{Shapovalova2010, Grier2013}.

On the other hand, the presence of observational signatures of inflow/outflow \citep[e.g.][]{Brotherton1994, done1996} makes it puzzling where this material comes from. Reasonable possibilities for the origin of the material are the AD and the interstellar medium. 
Recently confirmed flattened geometry of BLR resolved observationally for the first time in two sources \citep{GravityColl2018,GRAVITY2021} may favor the first possibility, i.e. the material originates from the AD; and also recent observational-based studies \citep{ribas2019a, ribas2019b, leftley2019} show that radiation pressure, likely on the multi-phase medium, is the prevalent mechanism responsible for the outflows in quasars.

There have been many attempts by different researchers to explain the formation and dynamics of BLR clouds in theoretical context over last 50 years.
The first attempt was to simply model the emission line formation by studying the photoionization equilibrium \citep{Osterbrock1978} which did not explain the BLR formation and dynamics \citep{Capriotti1980}.
The idea of magnetically-driven winds was then initiated by \cite{BlandfordPayne1982} in which the magneto-centrifugal force was responsible for driving the outflow to explain the BLR formation. Although the mechanism requiring a strong large-scale poloidal component of magnetic field was not self-consistent \citep{BalbusHawley1991}, dusty molecular torus and X-ray absorption features could be explained \citep{Elitzur2006, Fukumura2015,Huang2020}.
It was then proposed by \cite{Begelman1983} that the outer disk being irradiated by central source to the Inverse Compton temperature can give rise to formation of thermally-driven winds as the origin of BLR. This also required additional elements to be assumed such as hot corona above the disk or an inner hot flow \citep{Witt1997, BlandfordBegelman1999}. 
Radiatively line-driven winds as a mechanism for BLR formation firstly studied by \citep{murray1995} was the next theoretically-based attempt which was particularly/exceptionally successful in prediction of line profile shapes. This model most likely explains the HIL part of BLR emission, where line profiles show signatures of outflow. However, even line-driven wind does not always escape, and failed wind may form at some range of radii \citep{Risaliti2010}, thus complicating the line profile predictions.
Inflow models \citep{done1996} followed by others e.g. \citep{Hu2008, Wang2017} in which the flow of material from large radii was considered to be responsible for BLR formation also came under attention but they lack of predictive power.
BLR formation due to disk instabilities, also known as formation in-situ, has also been under attention for years \citep{Collin1999, Collin2008, Wang2011, Wang2012a}.

Radiatively dust-driven models are the newest models for BLR formation focusing on the radiation pressure of AD, and the presence of dust at large radii hinted in a number of papers \citep{Rees1969, Rieke1978, barvainis1987, Dong2008}. This dust is usually considered as constituting the dusty/molecular torus but dust can exist also closer in, if shielded, and may play a role in BLR formation. The models in this category are the Failed Radiatively Accelerated Dusty Outflow (FRADO) model \citep{Czerny2011}, and a model by \citet{baskin2018} in which the static disk is inflated due to irradiation in the dusty part (hereafter BL18).
In this work, we intend to study the dynamics of BLR based on the FRADO model.  Here we concentrate on the semi-analytical description of the cloud motion which will later allow to calculate the the BLR transfer function and line shapes, and to compare these predictions parametric models assuming arbitrary cloud distribution. In this way we aim at testing if FRADO is a viable model of the LIL part of BLR.

Due to various motivations like estimating the central BH mass, simulating the dynamics of BLR material due to radiation pressure, either in lines or on dust, has been of high interest for researchers \citep[e.g.][]{Marconi2008, NetzerMarziani2010, Wang2012b, Plewa2013, Shadmehri2015, Khajenabi2016}. Different geometries, and even time-dependent case of the radiation field has been investigated in the existing literature \citep[e.g.][]{Saslaw1978, Mioc1992, Liu2011, Krause2011}.
Some studies based on radiation pressure including hydrodynamics (HD) ones do not use a realistic prescription for the radiation field of AD. Either the disk has been initially considered as a point-like source with a luminosity \citep[e.g.][]{Donea2002, Mosallanezhad2019}, or it is treated as an extended object but only the radial component of the force (in spherical coordinates) is used in the computations \citep[e.g.][]{Risaliti2010}. Both cases yield the radiation field to be proportional to the inverse square of the distance so the radiation field is eventually in the form of a repelling central force never giving a strong vertical component close to the disk surface to lift the material sufficiently before being repelled outward. This can be applicable if BLR extends far beyond the outer radius of the AD, e.g. BLR models based on cloud inflow from larger radii \citep{Wang2017}, but this scenario is not firmly established. Instead, there are arguments that BLR is rather overlapping with the AD \citep[e.g.][]{kura2004}. There are also number of studies \citep[e.g.][]{proga1998, proga2000, proga2004, Nomura2020, Mizumoto2021} addressing the full disk radiation field but focusing on line-driving force in which the element of dust is missing.

Intuitively, the radiation force acting on clouds in the vicinity of AD is not radial but almost plane-parallel. This case has been investigated in FRADO model analytically but only in one dimension and for a very local flux \cite{Czerny2015,Czerny2017}. This 1-D model roughly reproduces the overall BLR position without the arbitrary parameters, but the line shapes based on this local model always showed traces of the double-peak structure, the vertical acceleration was not as efficient as required, and they did not show the characteristic Lorentzian wings observed in high Eddington ratio sources. The line shapes improved when much higher wavelength-averaged dust opacity was used than the standard $\kappa = 8$ [cm$^2$/g], but it was not clear that such opacity requirements were realistic. Thus, the issue of the radiation field of an extended disk must be addressed for which the net radiation force is a complex function of cloud position. Reaching to a realistic prescription for the radiation of AGN and the resulting outflow will be of high importance not only particularly in the physics of BLR but also generally in AGN feedback and star formation in host galaxies \citep{Croton2006, Raouf2019}.

Radiation field of an extended luminous disk was first calculated by \citet{icke1980} in a wavelength-averaged form. The method was only then followed in some papers \citep{Pereyra1997, proga1998, proga1999, proga2000, proga2004, feldmeier1999, Watarai1999, Hirai2001, Nomura2020}.
We present a detailed note on calculation of radiation field of an extended disk in the appendix of this paper and in a very generic form capable of being folded with wavelength-dependant dust opacities.

Although the proposed theoretical attempts mentioned above could somehow explain the formation of BLR along with some features of AGNs spectrum, they have not been able to self-consistently explain dynamical behavior of BLR and its geometry altogether which we have from observational data especially RM-based data. Hence, approaching to a unified self-consistent theoretical-based model which can explain the formation, geometry, and dynamics of BLR is highly required. The 3-D FRADO model is expected to recover observationally known features of BLR, i.e. radius or equivalently time-delay, complex velocity field, line-profiles, all self-consistently without the need of any arbitrary parameter(s). In this work we especially focus on the dynamics and geometry of BLR as the first step.

We therefore introduce our 3-D FRADO model in detail in section \ref{sec:realmodel}. The model does not go through HD calculations so the effects of the pressure gradients are neglected. However based on our results provided in section \ref{sec:results} most of the time clouds (assumed being pressure-confined) move at supersonic velocities (likely along with the surrounding medium due to dynamical coupling), so we preferably limit ourselves to the relatively simple and computationally efficient non-HD version of our model. We will then test in the next paper how much the model successfully catches the overall dynamical properties and explains the observational properties of the BLR, like line shapes, average time delays and and transfer functions for LIL lines like H$\beta$ and Mg II. The results are then followed by a discussion in the last section \ref{sec:discussion} where the advantages of our approach over other available BLR models including parametric models and hydro models are also addressed.

\section{3-D FRADO model with wavelength-dependent opacity} \label{sec:realmodel}

The basic FRADO model \citep{Czerny2011, Czerny2015, Czerny2016, Czerny2017} described the failed dusty wind motion in 1-D approximation, i.e. the motion vertical to the disk plane was included, apart from the rotational velocity. In addition, the wavelength-averaged dust opacity has been used. The radiation from the central region was only included in computations of the dust evaporation but neglected in the dynamics which was determined only by the local flux of AD. This approximation was necessary to formulate a semi-analytical model of the dynamics. However, the radiation pressure from the entire disk in this model is not properly represented, and the proper description of the dust interaction with the radiation field is also critical for the model.

In this paper we formulate the full 3-D model describing the motion of the initially dusty clouds under the radiation pressure coming from the entire AD and gravity of a central supermassive black hole (BH). We assume that the mass of the AD is much smaller than the black hole mass and its gravitational field is negligible. The net radiation force is neither radial nor vertical but it is a complex combination of both, and the clouds before being accelerated are in a circular motion along a local Keplerian orbit at the disk surface layers so the resulting motion of a cloud forms a complex 3-D trajectory. The net acceleration is
\begin{equation}
\textbf{a}^{\mathrm{net}} =
\textbf{a}^{\mathrm{gra}} +
\textbf{a}^{\mathrm{rad}}(<T_{\mathrm{s}})
\end{equation}
where $\textbf{a}^{\mathrm{gra}}$ stands for gravitational field of the central BH, and $\textbf{a}^{\mathrm{rad}}$ is the acceleration due to radiation pressure of AD given in its general form (find the detailed version in the appendix: equation \ref{eq:rad_pres2}) by
\begin{equation}
\int_{\lambda_{i}}^{\lambda_{f}}
\int_{\textbf{S}}
f \left(
I_{\lambda}, K^{\mathrm{abs}}_{\lambda},
K^{\mathrm{sca}}_{\lambda}, \Psi, \textbf{r}, \cdot\cdot\cdot
\right) d\textbf{S}\ d\lambda
\end{equation}
where
$I_{\lambda}$ is the radiation intensity of the AD specified in the section \ref{sec:diskproperties};
$\Psi$, $K^{\mathrm{abs}}_{\lambda}$, and $K^{\mathrm{sca}}_{\lambda}$ are the dust-to-gas ratio, total absorption opacity, and total scattering opacity of the clump, respectively (see \ref{eq:K_value1}, \ref{eq:K_value2}, \ref{eq:K_value3}, and \ref{eq:dust-to-gas}), and $\lambda_{i}$ and $\lambda_{f}$ defines the range of effective wavelengths for an adopted dust model, all addressed in the section \ref{sec:dust}; $\textbf{r}$ is the position vector of the clump;
($<T_{\mathrm{s}}$) implies that the radiative acceleration is available as long as the dust temperature calculated in the cloud along the trajectory is below that of dust sublimation $T_{\mathrm{s}}$ otherwise dust evaporates and the cloud performs later just a ballistic motion in the gravitational field of the BH (section \ref{sec:dustsublimation});
and $\textbf{S}$ is the surface of AD or part of it as discussed in the section \ref{sec:shielding} on shielding effect.

\subsection{Properties of underlying AD-BH system } \label{sec:diskproperties}

We assume the source of the radiation field represented by the extended optically thick geometrically thin disk, described by the \citet{SS1973} standard model (SS73). The inner disk radius is set at $R_{\mathrm{in}} = 6 R_{g}$, the outer radius is located at $R_{\mathrm{out}} = 10^6 R_{g}$ in all models where $R_{g}$ is the gravitational radius of the BH defined as $R_g = G M_{\mathrm{BH}}/c^2$. The flux density is locally described as in SS73, including the no-torque inner boundary conditions
\begin{equation}\label{diskflux}
\mathbb{F} (R) = \frac{3G \dot M_{\mathrm{edd}}}{8 \pi} \frac{M_{\mathrm{BH}}\ \dot{m}}{R^3}
\left( 1- \sqrt{\frac{R_{\mathrm{in}}}{R}} \right),
\end{equation}
and the locally emitted flux is represented as a black body emission, i.e. the intensity is given by the Planck function
\begin{equation}
 I_{\lambda} = B_{\lambda} \left( T (R, \varphi) \right).
\end{equation}

The AD flux is thus described by the central BH mass, set as $M_{\mathrm{BH}}=10^{8} M_{\odot}$ corresponding to the mean value in quasars catalog of \citet{Shen2011} for which
\begin{equation}
R_{g} = 4.7867 \times 10^{-6} ~~ \mathrm{[pc]}
= 0.0057 ~~ \mathrm{[lt-day]}
\end{equation}
and the dimensionless accretion rate, $\dot m$, normalized to the Eddington value for the adopted $M_{\mathrm{BH}}$
\begin{equation}
\dot M_{\mathrm{edd}}= 1.399 \times 10^{26} ~~ \mathrm{[g/s]}.
\end{equation}

In this paper, we stick to three values of $\dot m = 0.01$, and $0.1$ (low Eddington ratios), and $\dot m=1$ (high Eddington ratio).

Description of SS73 for the flux of a non-relativistic radiatively efficient AD provides us with an azimuthally symmetric profile for the effective temperature of AD as a function of radius $R$ since
\begin{equation}
\mathbb{F} = \sigma_{\mathrm{SB}}\ T_{\mathrm{eff}}^4(R),
\end{equation}
where $\sigma_{\mathrm{SB}}$ is the Stefan–Boltzmann constant, so
\begin{equation}
T(R, \varphi) = T_{\mathrm{eff}}(R). 
\end{equation}

\subsection{Dust opacity} \label{sec:dust}

In general, the radiation pressure acting on a BLR cloud should be a combination of absorption in lines and absorption/scattering on dust. In FRADO model, aimed at modelling the LIL part of the BLR we assume that the dust processes dominate. In this case, differently from the line-driven winds in which Doppler effect is important, the absorption efficiency does not depend on the cloud velocity, and the equations given in Appendix~\ref{sect:appendix_A} include this assumption. However, the effect depends on the assumption of the dust chemical composition, grain sizes and the dust-to-gas mass ratio.

The value of $K_{\lambda}$ can be obtained using prescriptions from \cite{rollig2013} and \cite{Szczerba1997} for different dust models  with a given distribution of dust sorts and grain sizes. See equation \ref{eq:K_value3} for details on how to find $K_{\lambda}$ assuming a given dust model with a certain distribution.
We further assume the classical MRN dust model appropriate for the interstellar medium \citep{mathis1977} as also used in BL18 model.
This simple model consists of silicate and graphite grains.
Signatures of silicate in AGNs are seen \citep{netzer2007}; and the equilibrium temperature of the grains implies the presence of graphite \citep{clavel1989}. However amorphous carbon grains rather than graphite are most likely expected based on UV spectra \citep{Czerny2004}, and/or the grain size range is not typical \citep{gaskell2004} as the overall extinction curve is more similar to SMC curve than to the typical graphite-dominated Milky Way curve \citep[e.g.][]{richards2003,hopkins2004,zafar2015}. The actual content of the AGN dust is still under vigorous studies \citep{giang2020}, but we adopt here the classical approach.
The grain size distribution in MRN model is identical, and given by
\begin{equation}
dn_{i}(a) = n\ A_{i}\ a^{-3.5} \ da  , \qquad 0.005 \leq a \leq 0.25\ [\mu m]
\end{equation}
where $a$ is the grain size (radius), $i$ stands for the dust sort, and $A_{i}$ is the normalization constant determining the overall abundances of the grain sort which directly leads to a unique value for dust-to-gas mass ratio $\Psi$ (see equation \ref{eq:dust-to-gas}). For example, setting $A_{\mathrm{silicate}}=10^{-25.10}$, and $A_{\mathrm{graphite}}=10^{-25.13}$ \citep{WG2001b} in the MRN dust model one can obtain $\Psi = 0.00955$ \citep{rollig2013}. We set $\Psi=0.005$ thus adopting the mean value of Milky Way \citep{mathis1977}. Dust-to-gas ratio in quasars has not been yet sufficiently studied, so a firm value or range of values in this context is not yet available. However, a value of almost $0.008$ is estimated for an AGNs sample \citep{Shangguan2018}, and also a recent study by \citet{Jun2020} has addressed the dust-to-gas ratio in obscured quasars. $K_{\lambda}$ is independent of dust sorts abundances. The MRN dust model is taken for simplicity and realistic modelling of AGN dust content must be addressed in future studies.

Here we assume that the coupling between the dust and the gas is strong. The friction thus prevents dust to move faster than gas and momentum is transferred to the gas, so the dusty/gaseous cloud moves as a single entity. The coupling is generally quite effective in dense media, for example in star-forming molecular clouds \citep[e.g.][]{reissl2018, Hosseinirad2018}. The local gas number-density ($n_{\mathrm{H}}$)
of the BLR clouds is rather high, with most recent estimates for the LIL BLR of order of $10^{12}$ [cm$^{-3}$] \citep[e.g.][]{tek2016, panda2018, panda2019a, panda2019b, panda2020, panda2020a, panda2020b} so this assumption should be satisfied.

\subsubsection{Possibility of outflow launching in FRADO}

We first test if the AD surface can be a source of the wind outflow under the effect of the radiation pressure. \citet{Czerny2015} and \citet{Czerny2017} argued that in the upper part of the AD atmosphere the Planck mean opacity is much larger than the Rosseland mean opacity used in the disk interior so the disk is not considerably puffed up, and this larger Planck value, appropriate for wind driving, will lead to an outflow. However, BL18 postulated that the AD atmosphere, with dust included, will remain static. So first we check whether indeed the condition postulated by FRADO model is satisfied if the realistic description of opacity is included.

\begin{figure}[b]
	\centering
	\includegraphics[scale=0.63]{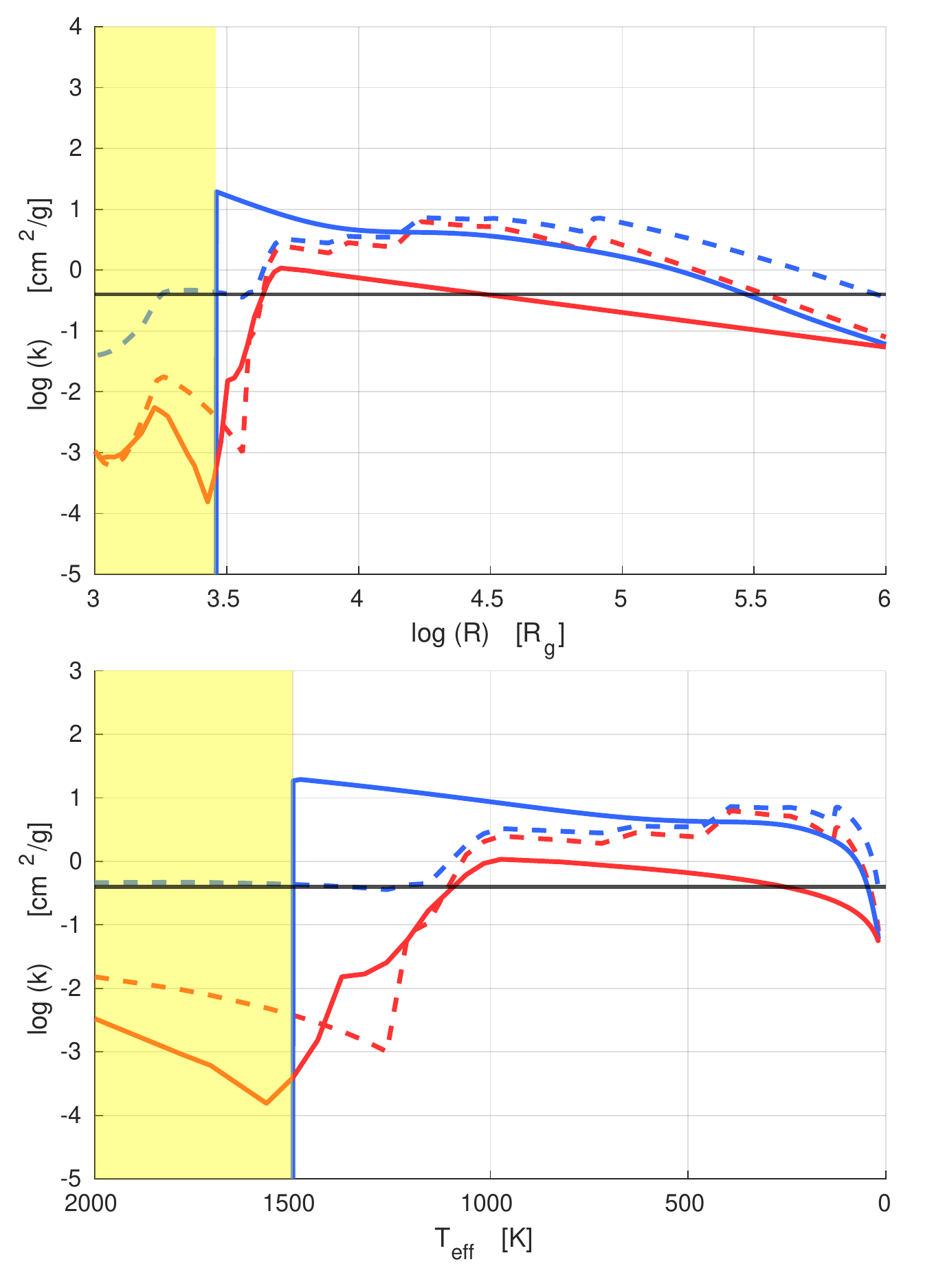}
	\caption{The upper panel shows the radial dependence of the Planck mean (blue solid-line) and the Rosseland mean (red solid-line) for $\dot m = 0.1$, $M_{\mathrm{BH}} = 10^8 M_{\odot}$ as used in the paper. The viscosity for calculation of Rosseland means is set to be $0.02$. The Planck (blue dashed-line) and Rosseland (red dashed-line) mean from the code of \citet{Semenov2003} for the density profile from the disk atmosphere compared to our values. The lower panel displays the values in terms of equivalent effective temperature of AD. The black solid lines in both panels show the certain value of Thompson cross section over the mass of proton. The regions shaded in yellow show the region where the temperature is above that of dust sublimation.}
	\label{fig:opacity}
\end{figure}

For this purpose, we calculated the value of the Planck mean opacity as a function of AD radius (or, equivalently, as a function of AD effective temperature) for the typical values of the model global parameters: $\dot m = 0.1$, $M_{\mathrm{BH}} = 10^8 M_{\odot}$ as shown in the figure~\ref{fig:opacity}. We integrated the wavelength-dependent dust opacity in the whole wavelengths range, from $\lambda_{i} = 6.1995$ [nm] till $\lambda_{f} = 1.2399$ [mm]. The aim was to compare it with the Rosseland mean which was used to determine the AD vertical structure needed to get the AD height.

The Rosseland mean was calculated for the same range of radii, at the position of the bottom of AD photosphere $\tau = 2/3$. This Rosseland mean, depending on density and temperature, comes from the tables which includes dust, molecules, atomic opacities and Thomson scattering (for details, see \citealt{rozanska1999}).

In our calculation of Planck mean, we multiply the result by $\Psi$ since due to gas-dust coupling the gas content of the clump hitchhikes with the dust. However as the gas contribution to the opacity is not included, the Planck mean formally shows a sudden drop to zero at the dust sublimation temperature. We see that the Planck mean is always higher than the Rosseland mean, which provides an argument in favor of the dusty wind launching from the AD surface.

Since our opacities are not most accurate, we check this result using the code\footnote{\url{https://www2.mpia-hd.mpg.de/~semenov/Opacities/opacities.html}} described in \citet{Semenov2003} and calculating the Planck and Rosseland mean for the dust and gas material. We used the gas number-density profile ($n_{\mathrm{H}}$) from the AD atmosphere, which varied from $1.37 \times 10^{14}$ [cm$^{-3}$] down to $3.88 \times 10^{9}$ [cm$^{-3}$] at the outer radius. These computations also imply that the Planck mean is higher than the Rosseland mean, and the idea of the dynamical character of the BLR is well supported.

We have to stress, however, that the use of Planck mean values of the opacity does not fully represent the driving force due to radiation pressure, as given by Equation~\ref{eq:rad_pres2}. In the actual computations we use the wavelength-dependent opacity and the force acting on the particle is calculated through folding this opacity with the radius-dependent and spectral-dependent radiation flux.

\subsection{Dust temperature}\label{sec:dustsublimation}

If the dust in the cloud is overheated, it evaporates rapidly. Therefore, at each cloud position we calculate the dust mean temperature by integrating the heat absorbed by the grains embedded in the cloud and comparing it to the total cooling of all the grains, independently from their size. 

Dust cools down by instantaneous re-emission of the absorbed radiation in the form of an isotropic blackbody radiation as long as the dust temperature is below that of sublimation so
\begin{equation}
Q_{\mathrm{abs}} = Q_{\mathrm{emit}}(<T_{\mathrm{s}}).
\end{equation}

Once $Q_{\mathrm{abs}} = Q_{\mathrm{emit}}(T_{\mathrm{s}})$
the dust content of the clump evaporates and the clump follows a subsequent ballistic motion falling back to disk surface. See the appendix (section \ref{sec:criterion}) for details. This is a certain simplification since the carefully calculated dust temperature depends on the grain type and grain size. However, since we have to perform the temperature computations for each point of each cloud trajectory, this much less time consuming approximation is justified.

We later assume the fixed specific value of the dust evaporation temperature of $1500$ [K] \citep[e.g.][]{barvainis1987, li2007, Figaredo2020, Huang2020}, the same for all species and sizes, and we assume that the process happens instantaneously. The evaporation is indeed fast, of order of 1 day (see BL18). Although the temperature of 1500 [K] is sort of mean value for all grain sizes and species, the assumption of a single temperature is much less justified \citep[BL18,][]{tazaki2020, Temple2020}, and has to be treated as a first approximation. However, in our dynamical model computations of the selective evaporation would be too time consuming, while it could have been done in BL18 model.

\subsection{Shielding effect}\label{sec:shielding}

According to radiatively-driven wind models, launching an efficient outflow in not possible if the launching region is not shielded from irradiation by the central source \citep[see e.g.][]{gallagher2007,proga2007, higginbottom2014}. This protection from intense central radiation, so called the shielding effect, has been firstly generally postulated by \citet{shlosman1985}, and later by \citet{Voit1992} to justify the presence of Polycyclic Aromatic Hydrocarbons (PAHs) in AGNs.  It was firstly modelled by \citet{murray1995} where authors assumed a bulk of dense gas at the inner edge of the line-wind launching region blocking soft X-rays.

The shielding can be naturally caused by other clouds \citep{Kartje1999}; the wind itself because of high-ionization parameter and high column density at the inner edge of the wind known as ``warm absorber'' \citep{MurrayChiang1995}, a magnetocentrifugal wind \citep{everett2005}; the innermost failed winds first found in hydrodynamical simulations by \citet{proga2000}; or the disk itself \citep{wang_disk2014}. Thus the apparently necessary shielding, particularly close to the disk surface, protects the wind medium from becoming over-ionized and likely prevents the radiation from the central parts to reach the distant disk regions \citep{Miniutti2013}.

There are many studies hinting the importance of and modelling the shielding effect \citep[see e.g.][]{proga2004, Risaliti2010, Sim2010, Higginbottom2013, Nomura2013, Hagino2015, Mizumoto2019}.

In order to incorporate the shielding effect into our model, or indeed to mimic the physical action of the complex multi-phase surrounding, we introduce two simple geometries. In our preliminary study of the shielding effect \citep{Naddaf2020} we have found that these two models are conveniently catching the required properties. However, in this paper we reintroduce them with some changes in their names and geometrical properties to make them mathematically and intuitively more intelligible.

\subsubsection{$\alpha$-patch model}

\begin{figure}
	\centering
	\includegraphics[scale=0.44]{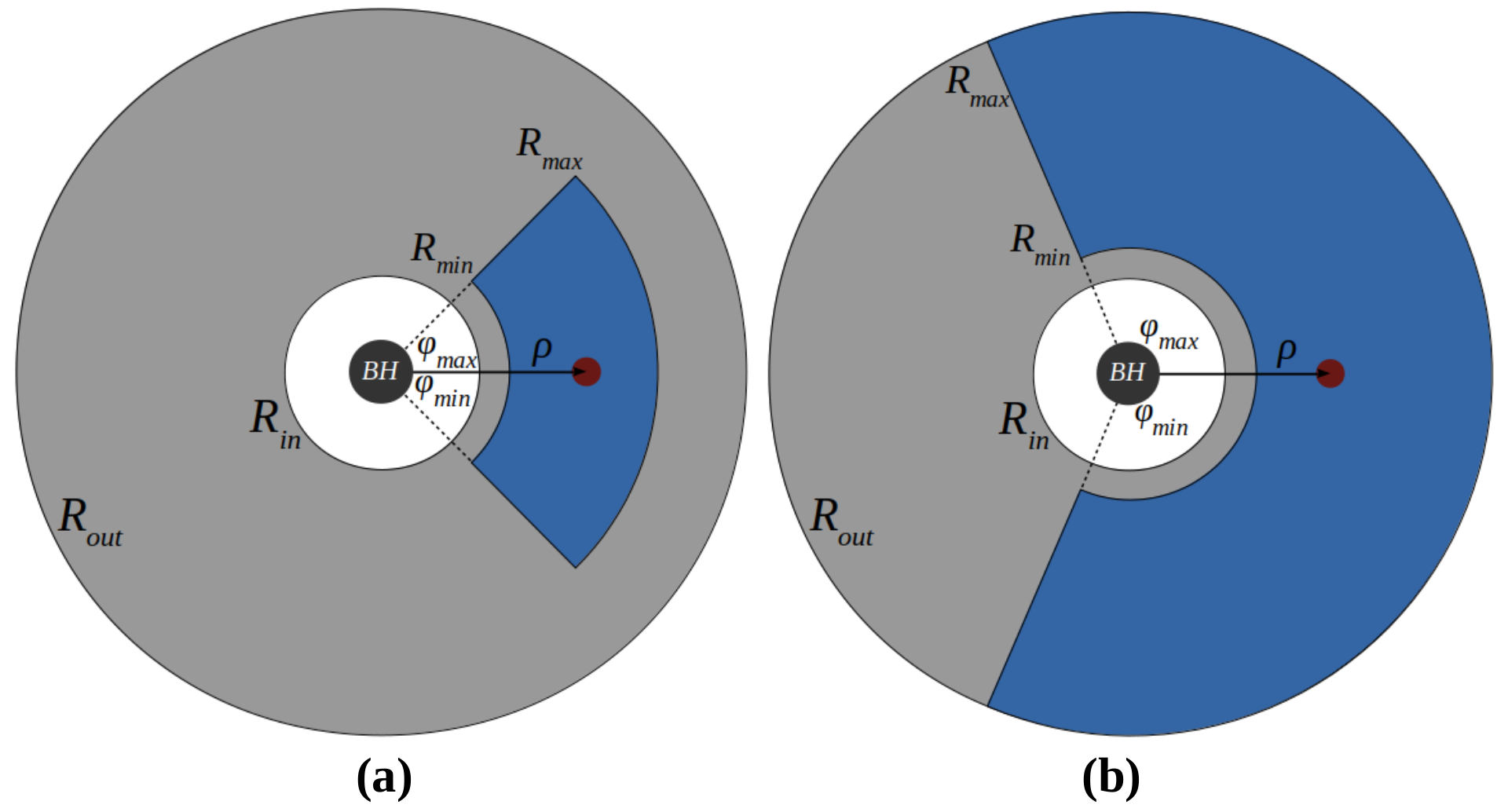}
	\caption{Geometry of (a) $\alpha$-patch and (b) $\beta$-patch shielding models. The cloud is exposed to a part of the disk radiation marked in blue. The size and location of these blue regions change as the cloud moves. The integral of radiation pressure to calculate the radiative force is integrated over the area of the disk shaded in blue, depending on the shielding model adopted.}
	\label{fig:SHgeometery}
\end{figure}

As the simplest form of shielding, we consider the contribution of radiation pressure from a small polar patch to act on the clump. We assume here that the cloud is exposed only to the radiation from the small part of the disk, centered below the cloud position, and the size of the patch is always proportional to the actual height of the cloud. Thus the position and the size of the patch varies as the cloud moves. A cloud very high above the disk is exposed to a large fraction of the disk. This approach mimics the fact that a given cloud is actually embedded in the rising clumpy wind, and this clumpy wind is much denser close to the disk surface, so the radiation cannot easily penetrate the medium in the direction roughly horizontal to the disk plane. This geometry is illustrated in figure \ref{fig:SHgeometery}(a). As a model parameter, $\alpha$, we introduce the ratio of the cloud height to the patch size, which is fixed during the cloud motion.

In order to incorporate this model into our computations, we apply the below upper and lower limits in the calculation of integral of radiative force
\begin{equation}\label{eq:alphapatch}
\begin{array}{ll}
R_{\mathrm{min}} & = \rho - \alpha z  \\
R_{\mathrm{max}} & = \rho + \alpha z  \\
\varphi_{\mathrm{min}} & = - \pi \left( \frac{\alpha z}{\rho - R_{\mathrm{in}}}\right) \\
\varphi_{\mathrm{max}} & =  \pi \left( \frac{\alpha z}{\rho - R_{\mathrm{in}}}\right)
\end{array}
\end{equation}.

\subsubsection{$\beta$-patch model}

In this model of shielding, the contribution of radiation pressure from a wider field of view is considered compared to $\alpha$-patch model. Although we still have a polar patch acting on the clump, we examine how considering a wider azimuthal angle and also outer part of the AD affects the motion of clumps and overall shape of BLR. We intend with this model to mimic the possible asymmetry of the shielding, i.e the fact that the outer part of the disk and the azimuthally extended region excluding the central disk may be visible to the cloud. In this model, the outer radius of the patch is always $R_{\mathrm{out}}$ of AD, the inner radius is proportional to the actual height of the cloud, and the patch is azimuthally extended by $\pi/2$. This geometry is shown in figure \ref{fig:SHgeometery}(b). The geometrical properties of the patch varies as the cloud moves.
So in this model again the increasing cloud height implies increasing exposure to the disk radiation.

Therefore, the upper and lower limits for the integral of radiative force are
\begin{equation}\label{eq:betapatch}
\begin{array}{ll}
R_{\mathrm{min}} & = \rho - \beta z  \\
R_{\mathrm{max}} & = R_{\mathrm{out}}  \\
\varphi_{\mathrm{min}} & =- \frac{\pi}{2} \left( 1 + \frac{\beta z}{\rho - R_{\mathrm{in}}}\right)  \\
\varphi_{\mathrm{max}} & =  \frac{\pi}{2} (1 + \frac{\beta z}{\rho - R_{\mathrm{in}}})
\end{array}
\end{equation}

The computation of the radiative force acting on the cloud presented in Appendix~\ref{sect:appendix_A} (equation \ref{eq:rad_pres2}) is general. If the shielding effect is assumed the integration is performed only for a fraction of the disk surface as set in equations \ref{eq:alphapatch} and \ref{eq:betapatch}. In order to neglect the shielding, one needs to set

\begin{equation}
\begin{array}{ll}
R_{\mathrm{min}} & = R_{\mathrm{in}} = 6\ R_{g}  \\
R_{\mathrm{max}} & = R_{\mathrm{out}} = 10^{6}\ R_{g}  \\
\varphi_{\mathrm{min}} & = 0 \\
\varphi_{\mathrm{max}} & = 2\pi
\end{array}
\end{equation}.

\subsection{Cloud dynamics}\label{sec:CloudDynamics}

In the present model we neglect the General Relativity effects and use just Newtonian dynamics since at the distance of the BLR these effects are usually relatively unimportant.

\begin{figure*}
	\centering
	\includegraphics[scale=0.7]{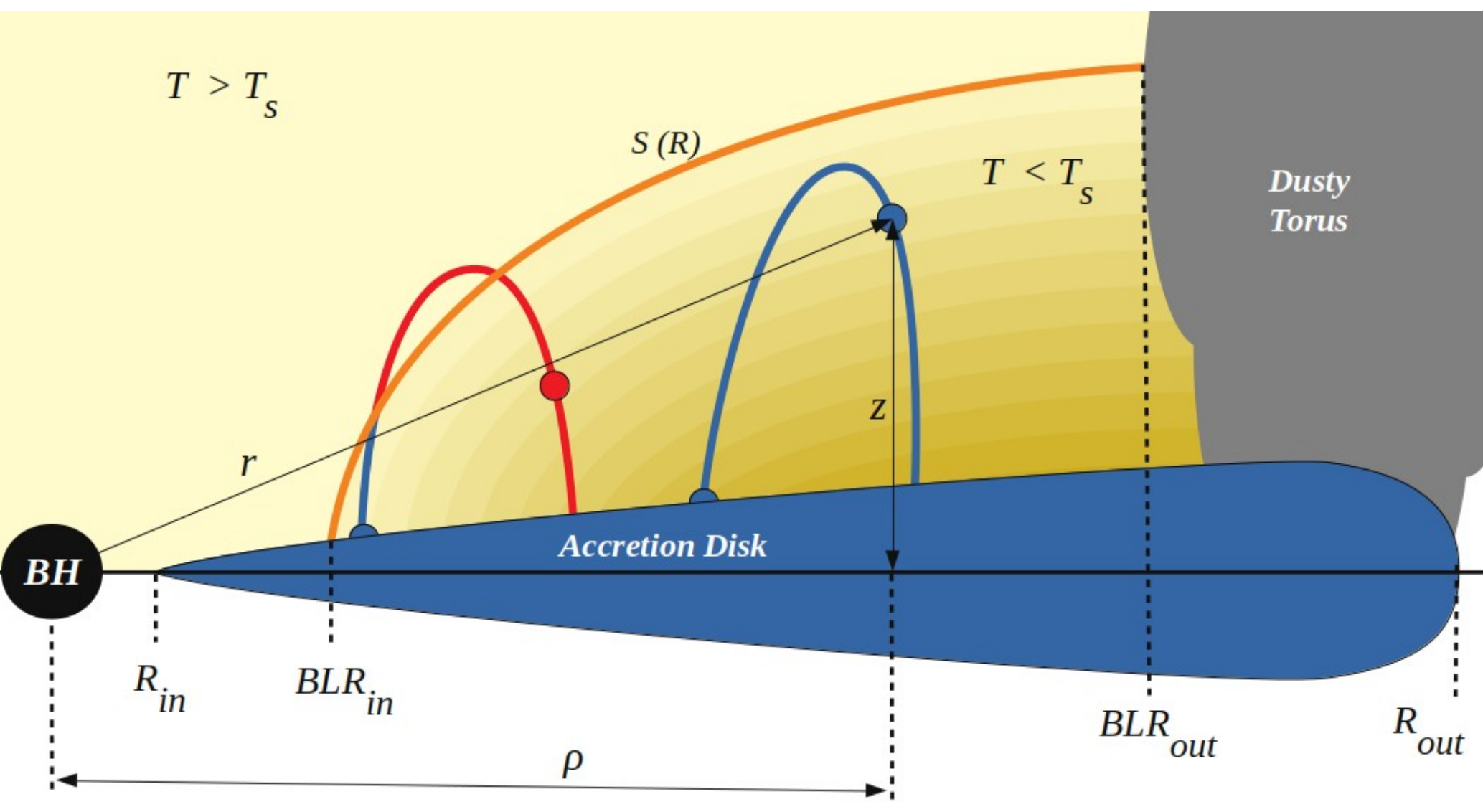}
	\caption{Schematic illustration of 3-D FRADO mechanism. The inner and outer radius of the AD are denoted by $\mathrm{R}_{\mathrm{in}}$ and $\mathrm{R}_{\mathrm{out}}$, respectively. The curved line in orange represents the \emph{sublimation location}, $S(R)$, above which marked by ($T>T_{\mathrm{s}}$) dust disappears due to intense radiation from the central parts of AD, called \emph{dust sublimation} region. The region restricted between disk surface and $S(R)$ marked by ($T<T_{\mathrm{s}}$) is called \emph{dust surviving} region where the temperature of dust is below that of sublimation. The onset of the BLR, $\mathrm{BLR}_{\mathrm{in}}$, is an innermost radius from which a cloud can be launched. $\mathrm{BLR}_{\mathrm{out}}$ sets the inner radius of the \emph{conceptual} torus, and the outer radius of BLR. A dusty clump being launched from the AD surface remains dusty (trajectory in blue) unless it reaches to $S(R)$ where it then loses the dust content and follows a ballistic motion (trajectory in red). Due to azimuthal symmetry, the position of the clump is described in the cylindrical coordinates. The radial position of the clump is denoted by $\rho$ to make is distinguishable from the integrand component of $R$ of the AD in the calculation of radiation pressure (see the appendix: figure \ref{fig:geometry_DiskCloud}(b), and equation \ref{eq:rad_pres2}). However, $\rho$ and $R$ are the same in nature, so for the matter of simplicity we use $R$ or $Radius$ in all plots. The vertical position of the clump denoted by $z$ is measured from the equatorial plane to which we refer by $z$ or $Height$ in all plots.}
	\label{fig:FRADO_mech}
\end{figure*}

The clouds are not allowed to cross the disk surface, so we also calculate the AD shape as a function of radius. We do that using the code from \citet{rozanska1999}, neglecting the effects of self-gravity on the disk structure \citep[see e.g.][and the references therein]{Czerny2016}. The disk thickness depends on the viscosity parameter which we set as $visc. = 0.02$ in our computations, motivated by the variability studies of AGN \citep[][and the references therein]{Grzedzielski2017}.

Clouds can be practically launched either from the disk surface or above it; the second approach illustrates the possible effects of the cloud collisions. But we only focus on launching from the disk surface (with zero vertical velocity). Clouds can be launched at arbitrarily large radius, but basically we concentrate at the launching radius lower than the outer disk radius so the clouds see the radiation field of the whole extended disk, and the radiation force - its direction and value - has to be calculated by integration over the disk surface. We consider models when the clouds see all the disk emission, but also we formulate models with shielding effect included. The inner radius for launching the clouds from the disk surface is actually the onset of BLR shown in figure \ref{fig:FRADO_mech} as $\mathrm{BLR}_{\mathrm{in}}$, or in other words the inner radius of the BLR for specified values of $M_{\mathrm{BH}}$ and $\dot m$ is set by the condition that a cloud can be launched.

\section{Results}\label{sec:results}

\begin{figure*}
	\centering
	\includegraphics[scale=0.5]{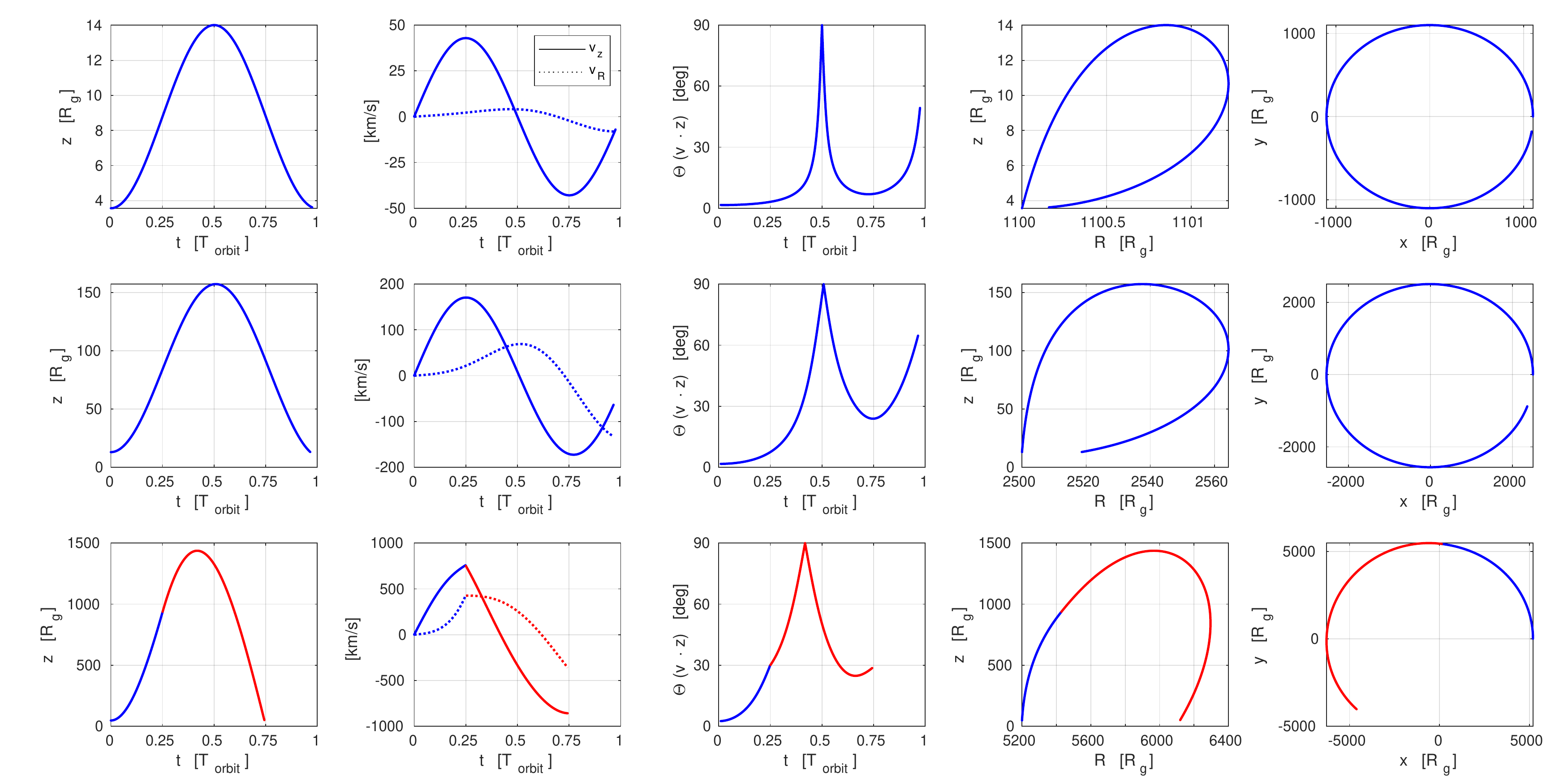}
	\caption{Example of motion for $\alpha$-patch model of $\alpha=3$. The results are shown for $\dot m$ = 0.01 (upper panels), 0.1 (middle panels), 1 (lower panels). The disk surface is not depicted. The motion of dusty cloud is shown in blue. Once the dust content of the cloud is sublimated, the subsequent ballistic motion is shown in red. Variables $x$ and $y$ are coordinates in the equatorial plane of AD so that $R^2 = x^2 + y^2$. The actual radial and vertical position and velocity of the model cloud in cylindrical coordinates denoted by $R$, $z$, $v_{R}$, and $v_{z}$, respectively. The temporal axis is denoted by $t$ in units of local Keplerian orbital time ($\mathrm{T}_{\mathrm{Orbit}}$). The middle column of panels shows the angle between the velocity vector of the cloud and the normal vector of equatorial plane in the frame locally co-moving with the disk. For these exemplary motions, an arbitrary launching radius of $R = 1100 R_g$, $2500 R_g$, and $5200 R_g$ are chosen for $\dot m$ = 0.01, 0.1, and 1, respectively.}
	\label{fig:single_cloud}
\end{figure*}

We formulated 3-D version of the FRADO model of the BLR. Our model does not (almost) have arbitrary parameters, apart from the values of the BH mass, $M_{\mathrm{BH}}$ and the accretion rate, $\dot m$ in Eddington units. The basic remaining free parameter is the dust sublimation temperature, $T_{\mathrm{s}}$. These three values of $M_{\mathrm{BH}}$, $\dot m$, and $T_{\mathrm{s}}$ determine the geometrical properties of the BLR and the 3-D motion of the clouds. Disk height value needed for the vertical position of cloud launching depends on the viscosity parameter in SS73 model, and if the shielding is considered, one more parameter characterizes its geometry. In the present paper we concentrated on the tests of the dynamical aspects of the model and the resulting geometry of the BLR.

For this purpose, we have set the model parameters as below:

\begin{equation}
\begin{array}{ll}
M_{\mathrm{BH}} & = 10^8~ M_{\odot} \\
\dot m & = 1 ~~~~ \mathrm{(High~Eddington~rate)}  \\
\dot m & = 0.1~,~ 0.01 ~~~ \mathrm{(Low~Eddington~rates)}  \\
T_{\mathrm{s}} & = 1500 ~~ \mathrm{[K]} \\
visc. & = 0.02 \\
\alpha & = 3~,~ 5 \\
\beta & = 3~,~ 5 \\
\Psi & = 0.005 ~~~~ \mathrm{(MRN)}
\end{array}
\end{equation}.

\subsection{Sublimation location}

Calculating the geometrical location where the relation $Q_{\mathrm{abs}} = Q_{\mathrm{emit}}(T_{\mathrm{s}})$ holds, called \emph{sublimation location} denoted by $S(R)$, we divide the space above the disk into two regions of \emph{dust sublimation} and \emph{dust surviving} as shown in figure \ref{fig:FRADO_mech}. The crossing radius of $S(R)$ and disk surface yields the radius where dust gets sublimated at the disk surface known as $\mathrm{BLR}_{\mathrm{in}}$. It sets the inner radius for launching the clouds as shown in figure \ref{fig:FRADO_mech}. Theoretically, BLR material can be present up to a radius known as $\mathrm{BLR}_{\mathrm{out}}$, which is shown in figure \ref{fig:FRADO_mech} and defined by the condition that dust can survive irradiation by the entire disk if the radiation field is spherical.

Without inclusion of the shielding effect, the function $S(R)$ has a bowl-like form (when viewed in 3-D) \citep[e.g.][]{ Kawaguchi2010, Kawaguchi2011, Czerny2011, Goad2012, Oknyansky2015, Figaredo2020}, while including shielding turns its shape to be more funnel-like.

These two values of $\mathrm{BLR}_{\mathrm{in}}$ and $\mathrm{BLR}_{\mathrm{out}}$ set the radial domain of our computations of the motion and trajectories of clouds. We have previously computed them and the function of $S(R)$ as available in \citet{Naddaf2020} for different values of $\alpha$ (or $\beta$) and also for the case with no shielding. Note that due to our change in the mathematical illustration of shielding models, the previous $\beta$ behaves like $1 / \beta$.

\subsection{Examples of 3-D motion of a cloud}
\label{sect:3D}

When the clouds are initially located at the disk surface, dusty wind is not launched without a shielding effect, as the disk surface, without any protection, becomes over-heated.

However, if a shielding is postulated, the initial outflow is easily launched. In figure~\ref{fig:single_cloud} we show several examples of an individual cloud trajectories. The path covered by a dusty cloud is marked in blue, and in case the cloud was exposed enough to cause the dust sublimation, the corresponding part of the trajectory is plotted with the red line. This part of the motion is just a ballistic motion in the gravitational field of the central BH, without radiation pressure force. 
In the dustless falling clouds, the dust can not form again during ballistic motion till hitting the disk surface, due to short time-scale and low-pressure \citep{elvis2002}.
Clouds also perform an orbital motion, and they preserve the original angular momentum they had at the launching radius. We show different perspectives of their trajectories, where z-axis is perpendicular to the disk plane and, the axes x an y are in the plane. We see already from this plot that the character of the motion strongly depends on the Eddington ratio.

\begin{figure*}
	\centering
	\includegraphics[scale=0.61]{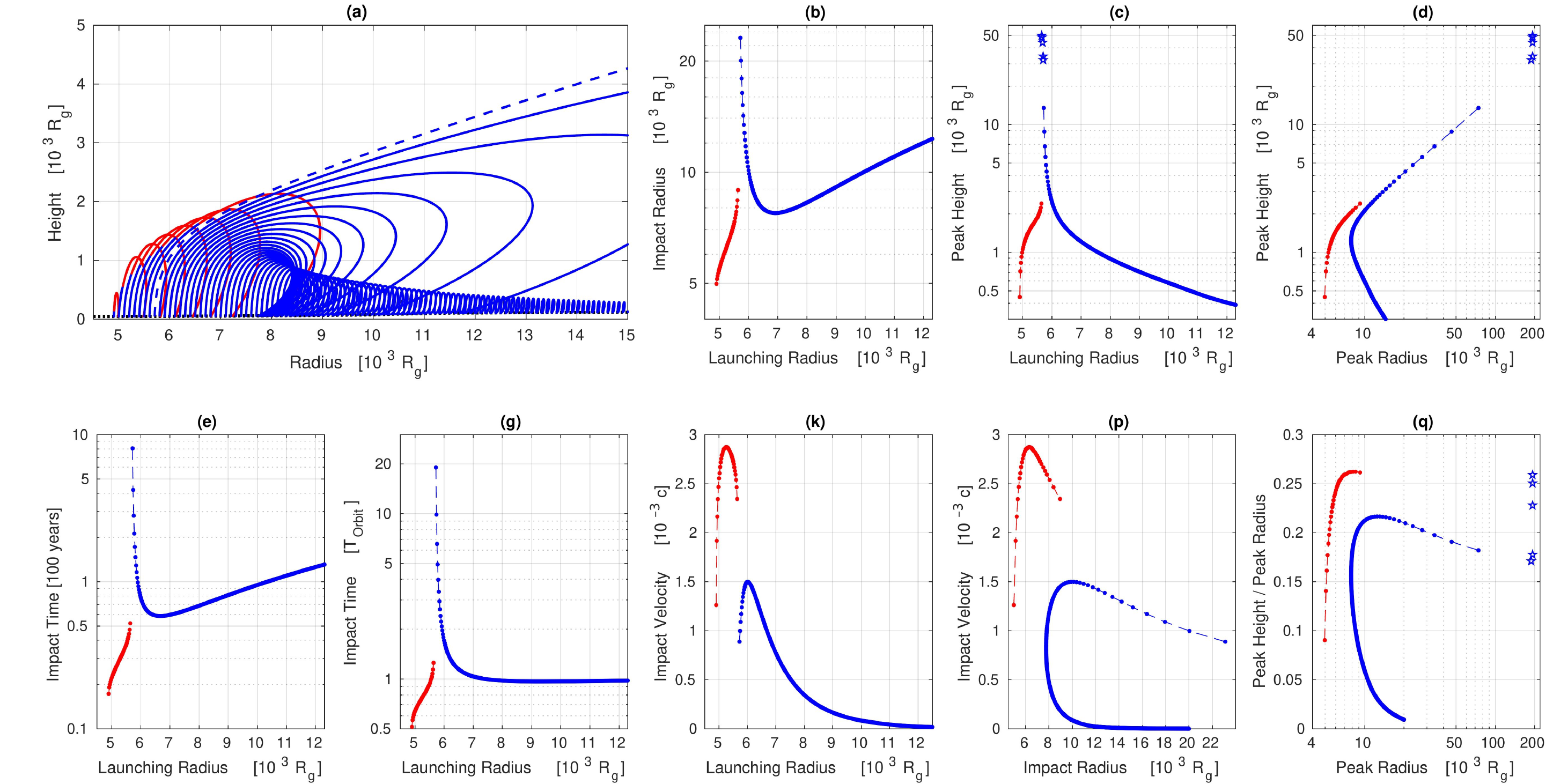}
	\caption{Global parameters of the cloud motion for $\dot m = 1 $, $\alpha$-patch model of $\alpha =3$. Escaping clouds are represented by asterisks.
	(a) The full trajectories of model clouds. Those trajectories in the inner radii starting in blue followed by red represent those clouds reaching to sublimation location where they lose their dust content and follow a subsequent ballistic motion. The trajectories fully in blue are the clouds which remain dust-full during their time-of-flight. The blue dashed-line represents the trajectory of a dusty escaping cloud.
	(b) The radii at which a launched cloud from a certain radius impacts the disk surface. As can be seen, there is a narrow radial range where the clouds launched within which can radially travel very long to finally impact the disk surface or even escape.
	(c) The maximum height a launched cloud from a certain radius can attain. For the same narrow radial range as previous sub-plot, clouds can reach to very high heights or even escape.
	(d) This can be considered as the shape of BLR if all clouds are at their highest vertical position. The Peak Radius means the radius at which the cloud is at the maximum height.
	(e) The time-of-flight of clouds in units of hundred years which shows the time-scale for the life-time of BLR clouds.
	(g) The time-of-flight of BLR clouds in units of local Keplerian orbital time ($\mathrm{T}_{\mathrm{Orbit}}$).
	(k) The velocity at which the clouds impact the disk surface. It shows the clouds launched from inner radii can severely disturb the disk once they hit it.
	(p) The impact velocity of clouds in terms of where they hit the disk surface. Obviously, the whole radial range from which a cloud can be effectively launched is hit by the falling-back clouds with impact velocities ranging almost within 300 up to 1000 [km/s].
	(q) The aspect ratio (Peak Height/Peak Radius) of clouds which sort of shows the opening angle of BLR clouds in terms of viewing from outside. The larger the aspect ratio and the smaller Peak Radius are, the easier the clouds are to be detected from outside.
	Note that in all sub-plots from (b) to (q) the actual situation of clouds are color-coded by blue and red for dusty and dustless, respectively.}
	\label{fig:single_global}
\end{figure*}

\subsubsection{Characteristics of motion at high Eddington rate}

To show the character of the cloud motion in more representative way, we plot the global parameters of the cloud motion for $\dot m = 1$ (high Eddington ratio) as displayed in figure~\ref{fig:single_global}. As before, we code with the red color the clouds which end their evolution as dustless.
Those subplots show the actual situation of clump in terms of being dusty or dustless at the time of hitting the disk surface or reaching the peak height. Being red does not mean they were dustless at the time of launching.
The majority of clouds complete the motion in less than the local Keplerian timescale and return to the disk, although at a different radius than the launching radius.

As can be seen from figure \ref{fig:single_global}(a), clouds launched in the outer part of the disk (more than $7 \times 10^3 R_g$) do not rise very high above the disk, and they fall back at the radius not much larger than their starting position. The maximum height is a strong decreasing function of the initial radius in the outer disk. The impact velocity in this region is below 300 [km/s], but nevertheless it is not negligible. The impact of the failed wind clouds creates additional mechanical heating at the disk surface, which was postulated by many authors unable to model the LIL part of the BLR by just radiative heating \citep[e.g.][]{joly1987, baldwin2004, panda2018, panda2020a}. 

Clouds launched at intermediate radii (almost within $6 - 7 \times 10^3 R_g$) complete very extended trajectories, they achieve a considerable height and large radii, but finally they return to much smaller radius again after performing a quasi-elliptic motion since their motion is limited by the conservation of the angular momentum set at the launching radius. The impact velocity of these clouds lies within $300-450$ [km/s].

Clouds launched at the inner region (almost within $5 - 6 \times 10^3 R_g$) are soon exposed to irradiation strong enough to cause the dust sublimation. Although their ballistic motion still brings them to relatively large heights, the radial extension of those orbits is not as large as of the orbits of the clouds launched at intermediate radii. The impact velocity of these clouds is the highest, of order of 1000 [km/s].

The most interesting orbits are those of clouds launched within a narrow range between the innermost region and the quasi-elliptic motion region, so called \emph{escaping zone}. For more details see section \ref{sec:escapingstream}.

\subsubsection{Comparison to low Eddington rates}

We see from the figure \ref{fig:single_global} that the character of the motion for high Eddington ratio depends critically on the launching radius, unlike the low Eddington ratios which show sort of a simple up and down motion (see figures \ref{fig:BLR_shape2}, and \ref{fig:BLR_shape3}).
Due to simplicity of the motion for low Eddington ratios, the results are provided in this paper only for the shielding parameters of $\alpha$ (or $\beta$) equal 3.
The figure \ref{fig:single_cloud} shows that the sublimation happens mostly at high Eddington ratio. This also can be seen in figures \ref{fig:BLR_shape2}, \ref{fig:BLR_shape3}, \ref{fig:BLR_shape1}, and \ref{fig:BLR_shape1_2}.

As can be seen from figures \ref{fig:single_cloud}, \ref{fig:BLR_shape2}, and \ref{fig:BLR_shape3} in the case of low Eddington ratios, the departure of the clouds from the disk surface and the radial extension of their orbit are not considerable. However, from figures \ref{fig:single_cloud} and \ref{fig:single_global}, for high Eddington ratio, the height reached and radial range covered by an individual cloud are a significant fraction of the launching radius, and the motion is far more complex. Unlike the cases of low Eddington ratios with very simple motion, the case of high Eddington ratio shows two interesting features due to the complex pattern of motion. They are the formation of an stream of escaping material and also the enhancement of accretion process at disk surface due to the complex profile of landing/launching radius, as described in the following. 

The apparently simple motion of the outer dusty clouds at high Eddington ratio, or all dusty clouds at low Eddington rates in general have an interesting aspect. Although these clouds do not show considerable radial motion their impact onto the disk surface happens at relatively high angle with respect to the disk (see third column of Figure~\ref{fig:single_cloud}). This is related to the fact that the vertical velocity before the impact is slowed down by the radiation pressure (see second column of the same plot), the angular momentum of the cloud is conserved, and the clouds complete almost one local Keplerian orbit, so the cloud takes a trajectory taking it close to the starting radius. Thus the impact of the clouds in all cases (not only for innermost clouds in high Eddington solution) happens at some grazing angle. This could contribute considerably to the development of the turbulence in the disk outer layers. It is not clear if such details are resolved in the numerical HD simulations since the scale height of this part of trajectory is quite small.

\subsubsection{Stream of escaping clouds}\label{sec:escapingstream}

The clouds launched within \emph{escaping zone} escape the radial domain of the computations, set as $\mathrm{BLR}_{\mathrm{out}}$, or the inner radius of the dusty torus by definition. More precisely, they do not return to the disk and directly go to torus. This stream of escaping material is relatively narrow, for the model with $\alpha = 3$ the width of \emph{escaping zone} is $\Delta R = 51\ R_g$, and the zone starts at the distance of $R_{\mathrm{stream}} = 5650\ R_g$. It shows a surprising similarity to the stream postulated by \citet{elvis2000} (hereafter EL00) purely empirical at the basis of direct observational arguments.
This is also similar to the fast escaping streams in non-HD models \citep[][]{Risaliti2010, Nomura2013} and HD ones \citep[][]{proga1998, proga1999, proga2000, proga2004, Sim2010, Nomura2020}.

Our ratio of the stream width to the distance is indeed very small, $\sim 0.01$ for the adopted parameters, but if the funnel-like structure is filled with the BLR clouds densely, and the cloud's gas number-density is $\sim 10^{12}$ [cm$^{-3}$], as argued in BL18 model and a number of studies \citep[e.g.][]{tek2016, panda2018, adhikari2019book}, the total column density $N_{\mathrm{H}} = n_{\mathrm{H}} \times \Delta R $ across the stream measured close to the disk surface could be as high as $7.5 \times 10^{26}$ [cm$^{-2}$]. Of course, this is only the firm upper limit since the stream content depends on the details of the outflow launching (see section \ref{sec:discussion} for more discussion).
Importantly, however it seems this can be an evidence that in the failed wind scenario the local radiation pressure of AD can support dusty clumps with very large column densities \citep{netzer2015}.
The asymptotic shape of the stream can be roughly described as a straight line in 2-D perspective in figures \ref{fig:BLR_shape1} and \ref{fig:BLR_shape1_2}. The inclination angle predicted by the model are around $\sim 74^{\circ}$ and $\sim 79^{\circ}$ for $\alpha$ (or $\beta$) equal 3 and 5, respectively. These are not much higher than that in EL00 with $\sim 60^{\circ}$. The line-driven wind models also frequently predict wind which is more focused towards the disk plane \citep[e.g.][]{higginbottom2014}. The lower part of the stream of material may not finally escape since it shows a decrease in vertical velocity at the outer radius of the computational grid, i.e. $\mathrm{BLR}_{\mathrm{out}}$ where we stop the computations. But the upper part of the stream totally escape since the vertical velocity of material close to $\mathrm{BLR}_{\mathrm{out}}$ remains constant and its vertical position monotonically increases. This happens due to the fact that we define the torus like a wall in the outer disk. So if the clouds with highly radially extended trajectories hit the wall (or, equivalently, crosses $\mathrm{BLR}_{\mathrm{out}}$), they are considered as the escaping stream. In other words, the condition is based the location of torus not the escape velocity of clouds. As can be seen from figures \ref{fig:BLR_shape1} and \ref{fig:BLR_shape1_2}, the inclination angle of the stream slightly increases with $\alpha$ (or $\beta$).

\subsubsection{Landing vs. launching radius}
\label{sect:land}

We notice that due to part of the radiation pressure coming from the innermost region the clouds are systematically pushed outwards. This effect is particularly strong in the case of high accretion rate.

We show the ratio of the landing to starting radius for an exemplary model in the figure \ref{fig:single_global}(b). This systematic return of the clouds to a disk but at larger radius has considerable consequences. First, the cloud motion is calculated assuming a conservation of angular momentum so the cloud returning to the disk has lower angular momentum than the Keplerian angular momentum at the landing point.
This newly arrived lower angular momentum material does not immediately accrete since it gets mixed up with the disk material at the impact but the net angular momentum at the disk surface (the clumps do not likely enter deeply into the disk) is lower than Keplerian so the accretion at the disk surface can get enhanced.

\begin{figure*}[hbp]
	\centering
	\includegraphics[scale=0.6]{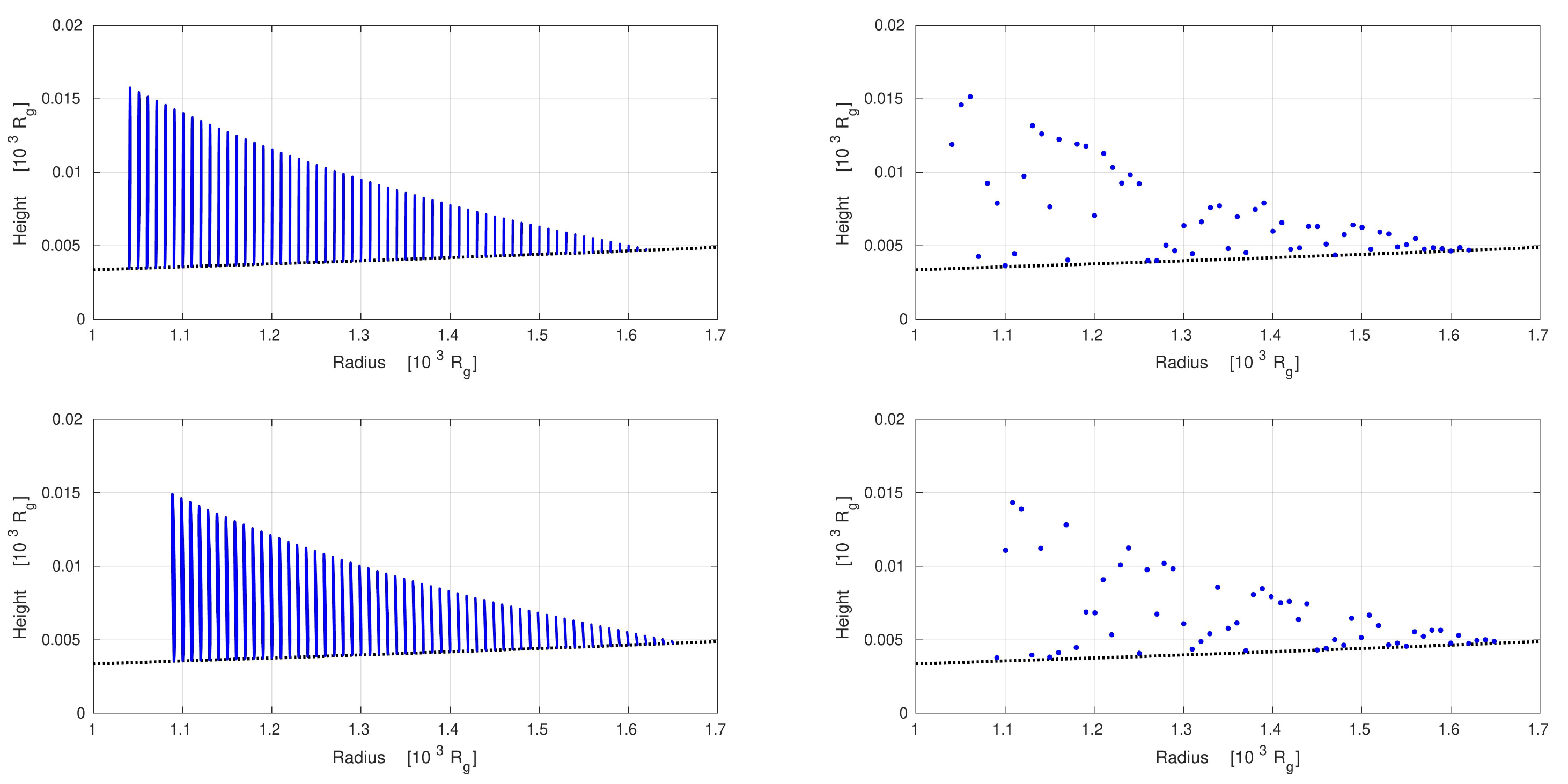}
	\caption{The trajectories of BLR clouds (left-panels) and shape of the BLR  in the form of a time-snapshot (right-panels) calculated from FRADO model for $\dot m = 0.01$. Upper panels: $\alpha$-patch model of $\alpha =3$. Lower panels: $\beta$-patch model of $\beta =3$. The black dotted line represents the disk surface.}
	\label{fig:BLR_shape2}
\end{figure*}

\begin{figure*}[hbp]
	\centering
	\includegraphics[scale=0.6]{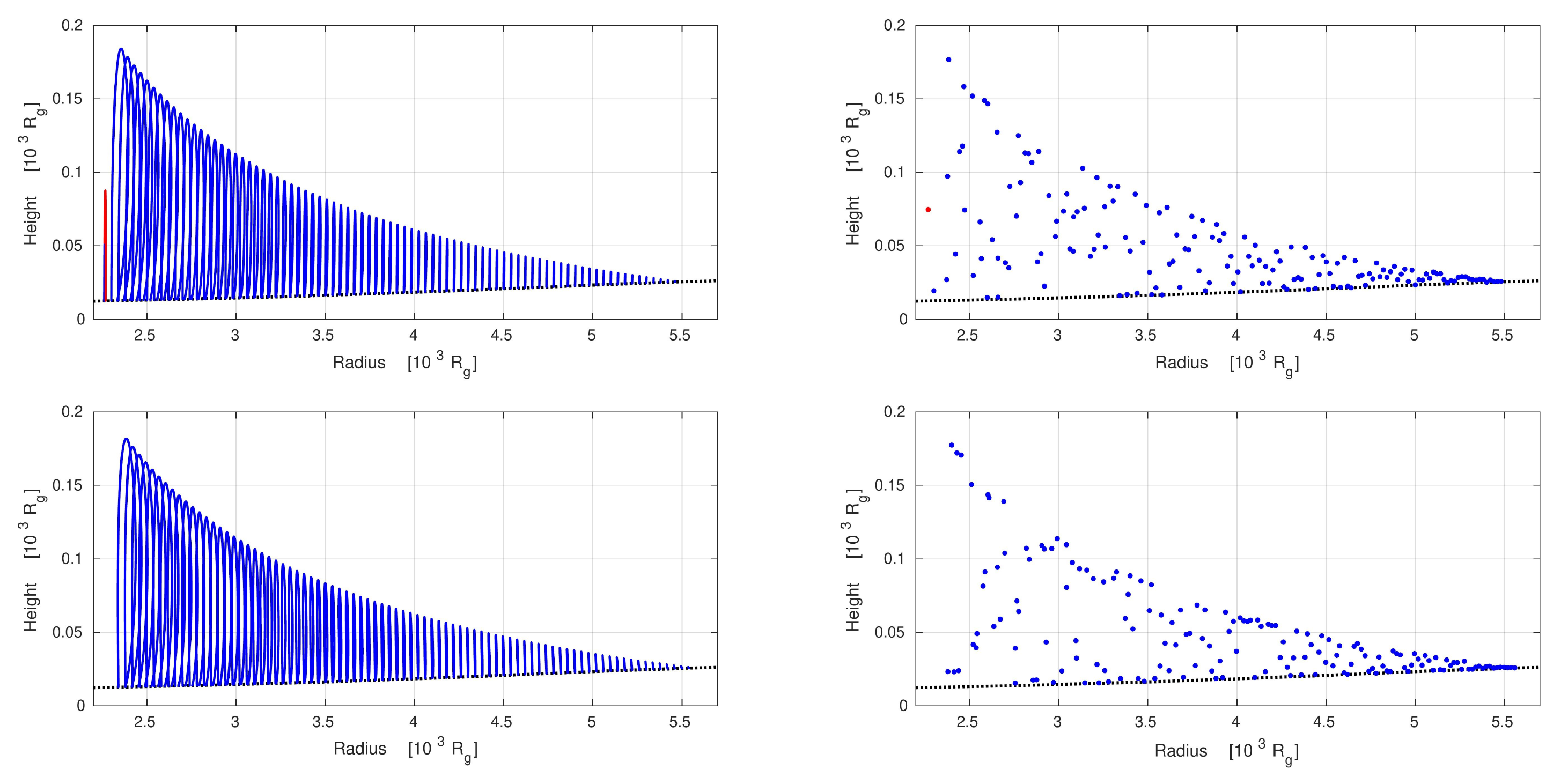}
	\caption{The trajectories of BLR clouds (left-panels) and shape of the BLR  in the form of a time-snapshot (right-panels) calculated from FRADO model for $\dot m = 0.1$. Upper panels: $\alpha$-patch model of $\alpha =3$. Lower panels: $\beta$-patch model of $\beta =3$.  The black dotted line represents the disk surface.}
	\label{fig:BLR_shape3}
\end{figure*}

\begin{figure*}[hbp]
	\centering
	\includegraphics[scale=0.5]{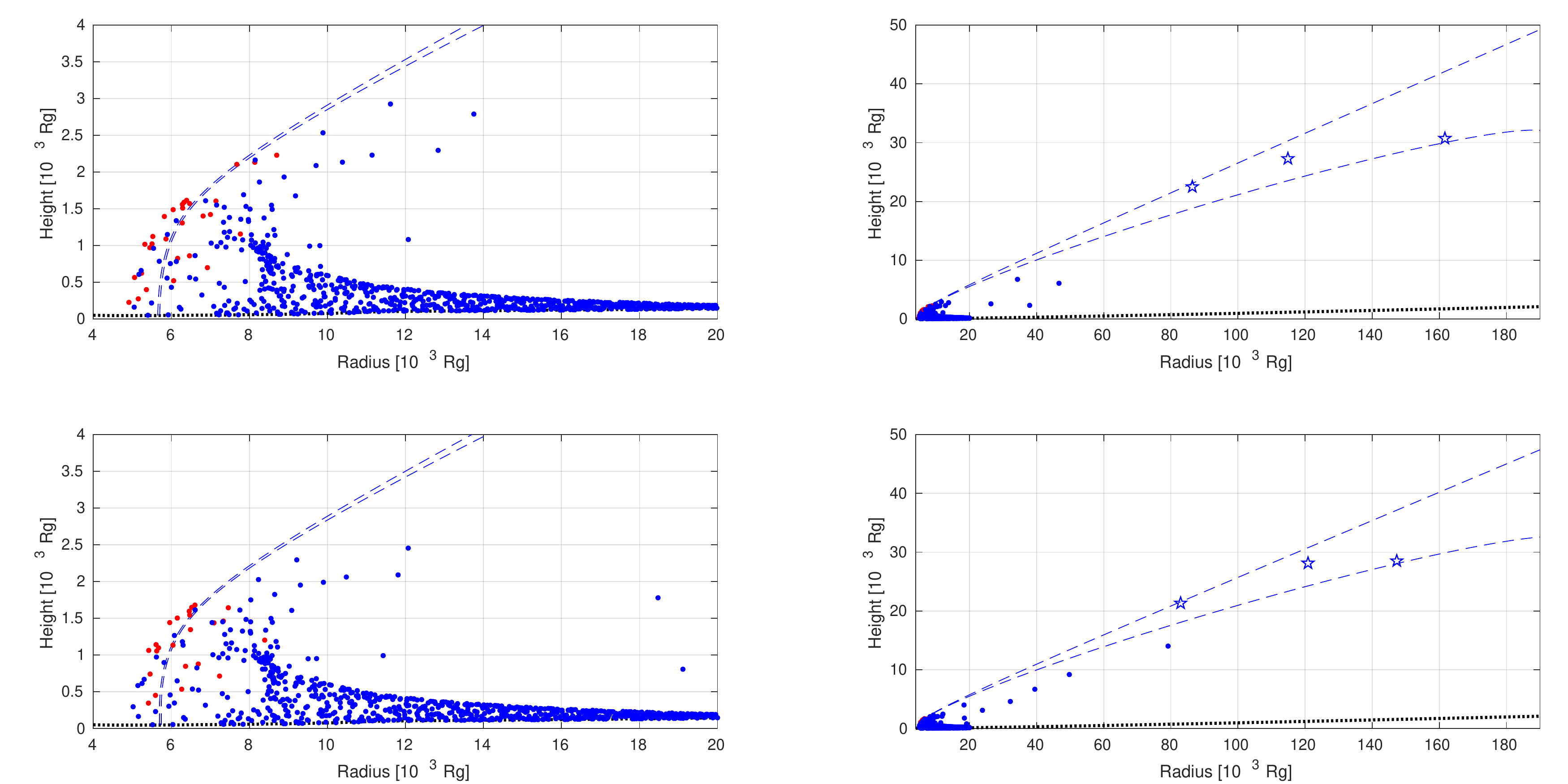}
	\caption{The shape of the BLR calculated from FRADO model for $\dot m = 1$ in the form of a time-snapshot. Upper panels: $\alpha$-patch model of $\alpha =3$. Lower panels: $\beta$-patch model of $\beta =3$. Right panels are the zoom-out version of left panels in order to show the extension of stream up to torus. The black dotted line represents the disk surface. Escaping clouds are represented by asterisks, and the area covered by escaping clouds is confined within two blue dashed lines.}
	\label{fig:BLR_shape1}
\end{figure*}

\begin{figure*}[hbp]
	\centering
	\includegraphics[scale=0.5]{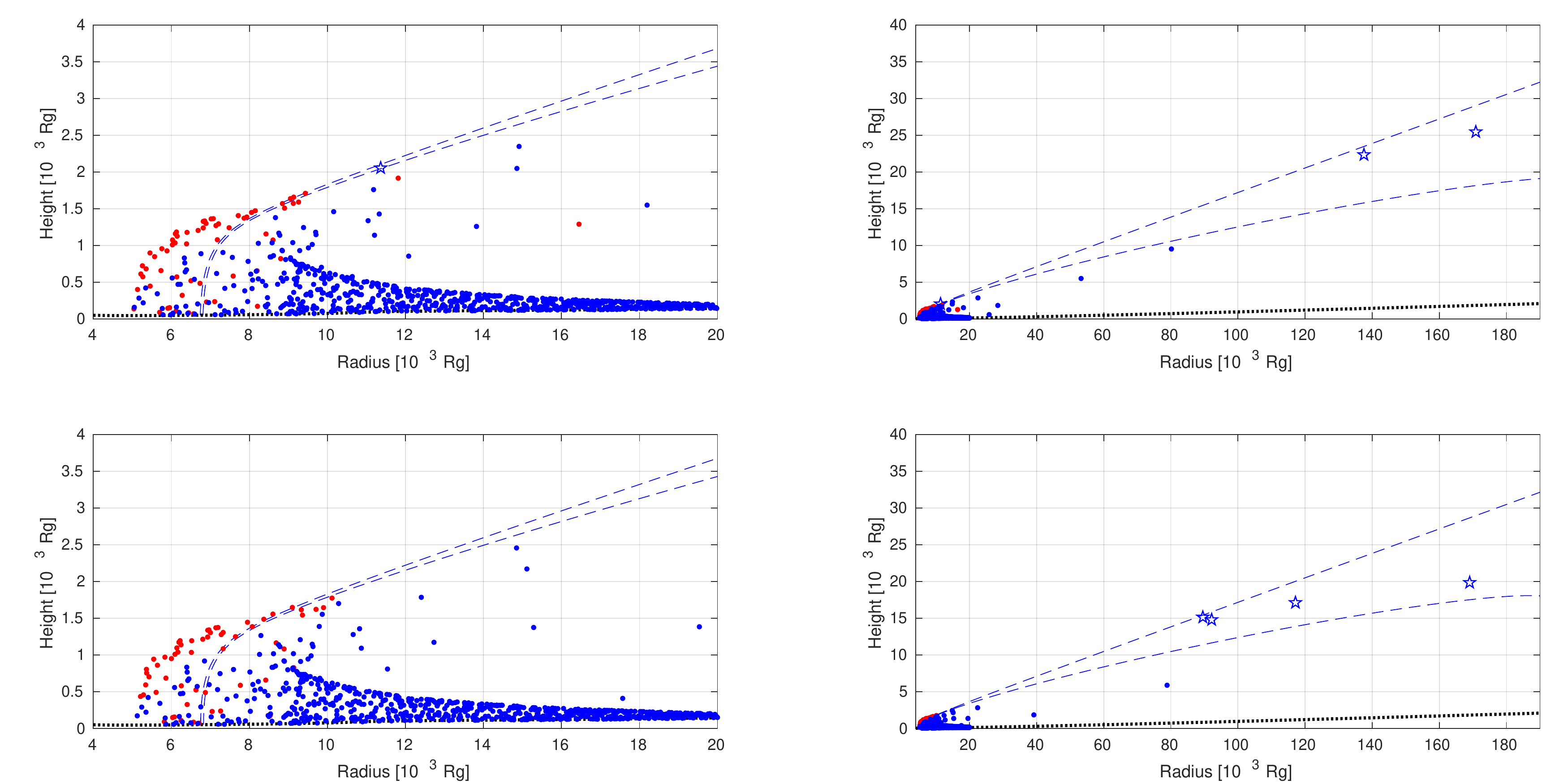}
	\caption{The shape of the BLR calculated from FRADO model for $\dot m = 1$ in the form of a time-snapshot. Upper panels: $\alpha$-patch model of $\alpha =5$. Lower panels: $\beta$-patch model of $\beta =5$. Right panels are the zoom-out version of left panels in order to show the extension of stream up to torus. The black dotted line represents the disk surface. Escaping clouds are represented by asterisks, and the area covered by escaping clouds is confined within two blue dashed lines.}
	\label{fig:BLR_shape1_2}
\end{figure*}

\subsection{The shape of the BLR}

In parametric models the BLR is usually represented as a cone filled with the clumps, and the radial and azimuthal cloud densities are assumed to follow a power law and exponential law, respectively \citep{Netzer1993, Ward2014, tek2016, Tek2018,GravityColl2018}.

In our model the shape is determined by the kinematics, and in particular by the maximum height achieved by any cloud for a given initial radius. This shape does not resemble a cone, it rises up relatively fast just after the onset of the BLR, and then the height becomes shallower. The net effect strongly depends on the accretion rate as well as the free parameters of shielding models.

In order to show the distribution of BLR clouds, we have taken a time-snapshot. For this purpose, a flat random number generator is used to choose a random position of a cloud during its flight. Each dots in the figures \ref{fig:BLR_shape2}, \ref{fig:BLR_shape3}, \ref{fig:BLR_shape1}, and \ref{fig:BLR_shape1_2} correspond to a moving cloud in BLR.

When accretion rate is low, the height of the BLR is not high, the clouds remain close to the disk surface, the Keplerian motion remains dominant, and the role of the vertical velocity is small (see figure~\ref{fig:BLR_shape2}, \ref{fig:BLR_shape3}). We see that for low accretion rates the BLR is close to the disk surface, and in practise such a model cannot be easily distinguished from the irradiated disk surface. Such a model is practically equivalent to BL18 model. In the present paper we do not yet perform the computations of the shape of emission lines but the simple consequence of small vertical height and small vertical velocity is the expected disk-like line profile. It is interesting to note that this trend is consistent with observational data: low accretion rate sources show frequently double-peak emission lines.

However, when the accretion rate is high, the vertical motion of the clouds is important. The aspect ratio (i.e. the ratio of the local BLR height to that radius) shown in figure~\ref{fig:single_global}(q) reaches 0.27 for dustless clouds, and 0.22 for dusty clouds. This maximum aspect ratio is not at the outer radius, but relatively close to the inner radius. We see from figures \ref{fig:BLR_shape1} and \ref{fig:BLR_shape1_2} that the overall shape, particularly for high accretion rate, is very complex. It does not resemble the simple shapes adopted in many numerical BLR models like disks, rings or shells \citep{pancoast2011}, or even more complicated but still regular patterns \citep[e.g.][]{pancoast2014}. It clearly does not support the previously mentioned conical geometry.
It is while for high accretion rate, however, a shallowing tail shown can be seen in figures \ref{fig:BLR_shape1} and \ref{fig:BLR_shape1_2} in the shape of BLR at the outer disk where the Keplerian motion is again dominant. This again resembles the results of BL18.

Overall, the calculated BLR shape in our model is similar to BL18. However, FRADO model can predict the missing part of the BLR shape in BL18 due to loss of static solution. Moreover, for high accretion rate the model also predicts the stream of escaping material proposed by EL00.

\section{Discussion}
\label{sec:discussion}

In this work aiming at developing a dynamical model based on the dust-driven outflow, we followed a non-HD semi-analytical approach to the physics of BLR. Compared to HD models the description of the dynamics is simplified through approximations described in section \ref{sec:approx3D}. Our motivation for this non-HD method is that it allows us to perform a reasonably fast test of the model for a wide range of initial parameters. This advantageous flexibility of our model versus HD models helps later to provide a large dense table grid which can be used for data fitting. Combining the results with line formation prescription in the next paper we will be able to compare it to the observational data (mean time delays, line shapes and transfer functions) for Low Ionization part of the BLR.

Our model, however simplified for the purpose of this test, relies on the relatively accurate description of the dust-driven dynamics, and later, when combined with outflow rates and line production efficiency, can provide us with a better insight into the BLR and distribution of material than purely parametric models, with density distribution of BLR clouds predetermined as an arbitrary function of the radius and distance from the equatorial plane \citep[e.g.][]{pancoast2014,Gravity2020}, although some of these models are quite complex \citep[e.g.][]{matthews2020}, involving conical outflow and a simple description of clumpiness. Such purely parametric models do not easily test the driving mechanisms. Our model does not have all the properties of the full hydrodynamical (HD) studies, however those studies are computationally very expensive. Moreover, any model before being implemented in HD context should be tested in an analytical/semi-analytical form. In the following subsections, we will discuss different aspects of the model, compare our results with other BLR studies including HD ones, and address the supporting observational facts.

In this work we calculated the 3-D structure of the BLR model within the frame of the FRADO model scenario. This dynamical model of the BLR is based on careful computations of the radiation pressure force acting on clouds launched from the disk surface. We included only the radiation pressure acting on dust, neglecting so far the line driving, but aiming at modelling the LIL part of BLR this approach may be a good approximation. On the other hand, we included the radiation force coming from extended disk (not the frequent assumption of a point-like source), and we included a description of the shielding necessary to launch the wind.

Our predicted shape of the BLR depends most significantly on the Eddington ratio. For high Eddington rate the material launched clouds are accelerated to high velocities at small launch radii, and part of them forms a stream of material which is outflowing. At larger radii, a simple failed wind is formed. This outflowing stream shows a striking similarity to the geometrical model of EL00 at the basis of observational constraints. At low Eddington ratios only a failed wind forms in the entire radial range, vertical velocities and clouds maximum heights are smaller. Despite the presence of a turbulence due to the clouds' impact velocities which is a sort of monotonically decreasing function of radius starting around 110 [km/s] (9 [km/s]) for $\dot m = 0.1$ ($\dot m = 0.01$) in the inner BLR down to less than 1 [km/s for] the last possible trajectory corresponding to last lifting radii, the BLR resembles somehow a static disk surface. Therefore it shows a considerable similarity to the static BL18 model, where clouds perform only an orbital motion while here clouds have vertical velocities.

The 3-D dynamical model is time consuming, so currently some processes are described in a simplified way. This can be improved later on, and here also we do not yet perform detailed comparison to the observational constraints.

\subsection{Approximations in the 3-D model}\label{sec:approx3D}

According to the column densities in BLR single cloud \citep{Bianchi2012} which are of order of $10^{23}$ [cm$^{-2}$], the clouds are moderately optically thick. This is also an assumption of our model since we use a single-scattering approximation (see Appendix \ref{sect:appendix_A}). If the clouds are optically thick the calculation of radiative transfer should be incorporated into our computations which is beyond the scope of this work. Notably, despite considering single-scattering approximation the model produces a powerful and fast outflow, contradicting the results by \citet{Costa2018} who argue extreme or unrealistic luminosity is required to launch an outflow if single-scattering is assumed. It is due to the fact that \citet{Costa2018} consider only the IR part of the AD spectrum, while in 3-D FRADO model broad range of frequencies from X-ray to UV to IR contributes to the radiation pressure acting on dust depending on the actual position of the dusty cloud.

Moreover, the clouds in our model are considered to be very compact i.e. almost point-like (see \ref{sect:appendix_A}), while they are loose extended objects \citep{Bianchi2012}.
We do not address the possible collisions between clouds, however later in \ref{sec:collisions} we show it can be a reasonable assumption. The friction with the ambient medium, and instability of clouds due to HD processes are also neglected. However we discuss our method of treatment of clouds and interactions with intercloud medium in \ref{sec:stability}.

Moreover, our description of the opacity is simpler than the one used in BL18 model. We assume a fixed single dust sublimation temperature, independent from the grain size and chemical composition. In reality, as in BL18 model, this temperature depends on the grain size, so the dust evaporation should proceed more slowly as the cloud moves, with smaller grains disappearing first, and large grains still providing some radiation force support to the cloud against gravity. This is not yet incorporated in our model.

\subsection{Shielding parameters}

We adopted arbitrary values for the two free parameters of $\alpha$ and $\beta$ in our shielding models. That was due to the fact that we only wanted to test and perceive an overall view of the dynamics of BLR material and its geometry in the 3-D FRADO model with sort of shielding effect included. Obviously from the figures \ref{fig:BLR_shape2}, \ref{fig:BLR_shape3},  \ref{fig:BLR_shape1}, and \ref{fig:BLR_shape1_2} the shape of BLR depends on the value of $\alpha$ (or $\beta$) parameter, although this dependence is very weak for low Eddington ratios. But the two presented shielding models, either $\alpha$-patch or $\beta$-patch being adopted, are not clearly distinguishable in terms of the resulted BLR shape. Taking a precise look at these figures, it can be seen that by adopting $\beta$-patch model, the overall shape of BLR shifts a bit outwards compared to $\alpha$-patch model. This implies that the wider azimuthal angle in $\beta$-patch model contributes to the total radiative force significantly, while the presence of outer radii of the disk in $\beta$-patch model is not of significant in terms of radiation pressure. So the contribution from large radii can be neglected in $\beta$-patch model as it is in $\alpha$-patch model. Thus the models of $\alpha$-patch and $\beta$-patch are slightly different only due to different range of azimuthal angles they cover. The $\alpha$ model, however, has an advantage over the $\beta$ model that it clearly shows the locally limited radiation pressure of AD is solely able to lift the material. The $\alpha$ model is consistent with the fact that the field of view of a cloud above the AD is limited due to presence of the ambient medium. The $\beta$ model takes into account the density gradient in the ambient medium, and may be an attractive alternative to $\alpha$ model when a detailed quantitative comparison with some observational data is done.
These results of the presented shielding models in this paper can help us to approach a realistic prescription for the shielding effect in a better physically-motivated way which we aim at in the next papers.

\subsection{Static vs. dynamical BLR model}

We stressed in the previous sections that our dynamical model for low Eddington ratios gives qualitatively similar results to BL18. However, some similarities with BL18 exists also for high Eddington ratio, only the basic approach of BL18 and FRADO is different. For high Eddington ratio, apart from the similar shallowing tail at the outer BLR, BL18 also notice a loss of hydrostatic equilibrium at small radii expecting formation of a wind where their static model does not fully work. The difference is that we follow the dynamical character of BLR under radiation pressure while BL18 only report the inability to obtain the static solution. Qualitatively, the loss of the static solution in BL18 happens at an almost similar location as the location of the \emph{escape zone} in our model, at about 0.03 [pc] (see figure 9 of BL18 and figure~\ref{fig:single_global}, \ref{fig:BLR_shape1}, and \ref{fig:BLR_shape1_2} in this paper). The quantitative comparison is not possible since BL18 use in their figure 9 the super-solar metallicity ($Z/Z_{\odot} = 5$) while we perform computations assuming the dust-to-gas ratio roughly corresponding to solar metallicity. The important comparison could be done in the future, when line profiles are calculated, and then the difference between the static model and the dynamic model would be likely quite clear.

\subsection{Cloud formation and their stability}\label{sec:stability}

Dynamical picture of the clumpy BLR model requires the physical justification of the assumed clumpiness of the medium. This is, however, a complex issue. Wind outflow from the disk is formally launched as a continuous medium, although in the case of a failed wind some level of clumpiness may appear already close to the disk surface. Stellar winds (escaping winds) develop clumpiness at some distance from the star \citep[e.g.][]{muijres2011}, most likely due to the development of the thermal instability in the irradiated medium. The expectations of the clumpiness were broadly discussed in various context  \citep[see e.g.][and the references therein]{mccourt2018,gronke2020}, and hydrodynamical simulations of this effect are difficult. In the context of AGN, spontaneous cloud formation from the wind in numerical simulations was reported by \citet{Proga2015, WatersProga2016, waters2019,waters2021} but the cold clumps falling down were found only in the last paper which studies much larger radii of accretion flow (well within the dusty torus), but still without including the dust content. So the situation in the context of the BLR clouds at expected distances is not clear.

As for the stability, the simple evaluation of the destruction (due to ablation or evaporation) of the clouds taking into account electron conduction and irradiation (M\"uller et al., in preparation) gives the timescales of order of 100 years, comparable to the orbital timescales. On the other hand, some very short episodes of X-ray absorption identified individual BLR clouds \citep{risaliti2011, torricelli2014}, most likely those exceptionally high above the disk surface so they were still along the line of sight towards the observer. The observational constraints for the clouds from eclipse \citep{pietrini2019} imply the clouds sizes of $10^{13}$ [cm] for a $10^8 M_{\odot}$ black hole mass. Other reports of the AGN clumpy medium like that by \citet{markowitz2014} contain a mixture of shorter events caused by the BLR clouds and longer events related to the dusty/molecular torus clouds, which complicates drawing a firm observational conclusions. Most of the clouds located very close to the disk are never seen in absorption since their presence (potentially visible in highly inclined sources) is hidden from the observer by a dusty torus.

Clouds are also unstable due to Kelvin-Helmholtz (KH) process for which a characteristic timescale can be estimated \citep[see e.g.][]{peisker2020}. The timescale for the growth of KH instabilities (for clouds of the size as in our model) is of order of few days (M\"uller et al., in preparation). This is much shorter than the flight time of clouds implying likely complete destruction of clouds before landing, however magnetized clouds can cope with KH instabilities \citep{McCourt2015} and survive longer \citep{shin2008}. In addition, there are also a number of HD simulations showing an efficient acceleration of dusty material and longer survival time despite the destructive role of such instabilities \citep[see e.g.][]{davis2014, tsang2015, zhang2017, zhang2018}.

\subsection{The effects of cloud-disk collisions}

Failed wind clouds returning to the disk bring several additional aspects which are not deeply discussed in the current paper. We  briefly discussed in Section~\ref{sect:land} landing issue in our model in terms of angular momentum. The impact radius of the cloud is always larger that the formation radius, which is due to the larger radiation flux from smaller disk radii. Since cloud preserve the angular momentum, in this way we have a systematic departure from the strictly Keplerian angular momentum at the disk surface due to the cloud impact. This may enhance the surface layer accretion since the effective angular momentum there becomes sub-Keplerian.

Next effect is due to the mechanical impact, considerable in the case of clouds on elongated orbits present in the $\dot m = 1$ solution, and less important for clouds with landing radius close to the starting radius. We can estimate the depth where the cloud material is deposited inside the disk by a simple comparison of the ram pressure with the gas pressure inside the disk, knowing the disk vertical structure, including the density, temperature and pressure profile. For models discussed in this paper and Eddington ratio of 0.001 the clouds are stopped at the height of order of 10-20\% lower than the total disk height, while for non-local clouds and Eddington ratio of 1.0 clouds graze at up to $\sim$50 \% of the total disk height. The impact leads to shock formation and a non-thermal emission from particles accelerated in these shocks (M\"uller et al, in preparation).  

The impact will lead to the destruction of the dust in the inner part of the BLR region, where the surface temperature of the unperturbed disk is below the dust sublimation temperature but clouds penetrating the disk reach layers where the temperature is higher. High velocity clouds impacting occasionally at larger radii can do that as well. The dust there is destroyed by the mechanical heating at the impact. However, the medium at the impact is rather dense, with the local number density $n_{\rm impact}$ of the plasma above $10^{15}$ [cm$^{-3}$], so it cools efficiently. If we use the cooling function $\Lambda(T)$ provided by \citet{gnat2012} we can see that the kinetic energy of impact, of order of $n_{cl} m_p v_{\rm impact}^2$ per unit mass can be re-emitted in a timescale of order of seconds
\begin{equation}
    t_{\rm cool} =
    \frac{ n_{cl} ~ m_p~ v^2_{\rm impact} }
    { \Lambda(T)~ n_{\rm impact}^2 } \sim 1~ [s]
\end{equation}
assuming the gas temperature $T$ of order of $10^4$ [K] and assuming that the cooling medium is still optically thin, as the whole cloud. The key issue is then the origin of the dust. If the dust must arrive from larger radii (e.g. from the outer dusty/molecular torus), then of course the accretion process would be very long. However, as we argued by \citet{elvis2002}, and used in the proposed FRADO model of  \citet{Czerny2011},  the conditions in the outer layers of the disk are perfect for the dust formation {\it in situ}. Efficient dust formation requires a well defined pressure and temperature conditions, and high densities in the disk allow to satisfy these conditions. The conditions of the dust formation are best studied in the case of stars, the details of the early stage of the dust formation are still unclear (the issue of dust seeds - see e.g. \citealp{ventura2012}) but the dust formation actually happens then in the stellar wind, at distances up to a few stellar radii, so the timescales must correspond to the wind outflow timescales, set by the wind velocity, of order of a few tens of [km/s] \citep[e.g.][]{Goldman2017}, which is of order of days up to a year. Such a timescale is much shorter than the time separation between the two cloud impact at a given location since the cloud impact events are separated by the timescale of order of a fraction of local orbital (Keplerian) timescale, which is about 100 years. Therefore, the dust destruction is only temporary, and dust content in the disk atmosphere is recovered between impacts.

\subsection{The fast outflow stream}

High velocities of escaping material which can extend to very large distances triggered by disk radiation pressure have been addressed in a number of papers \citep{Hopkins2010, Harrison2014, Ishibashi2015, Thompson2015, Ishibashi2017, Costa2020}.

The stream-like feature of the outflow predicted empirically by El00 was previously shown in HD simulations by \citet{proga1998, proga1999, proga2000}. More studies later also showed development of such feature in HD context \citep[see e.g.][]{proga2004, Sim2010, Nomura2020} and even in non-HD models \citep[][]{Risaliti2010, Nomura2013}.

While the models mentioned above focus on line-force on BLR gas, which most likely correspond to the HIL of the BLR, this work is the first study hinting for the development of such feature also at larger distance, in the LIL BLR (not torus), as a result of radiation pressure acting on dust. The stream is highly inclined and focused more toward AD with an inclination of $\sim$ 70 to 80 degrees, consistent with observation \citep{GravityColl2018} and sophisticated HD models \citep[e.g.][]{proga2000, proga2004, Nomura2020}. However, it initiates at larger radii of few $10^3 R_{g}$ responsible for the LIL BLR compared to few  $10^2 R_{g}$ in HD studies \citep[also see][]{Sim2010, higginbottom2014} where HIL BLR develops. Although the feature in our model is seen only for the case of high Eddington rate but an enhancement of metallicity and dust-to-gas ratio to more typical super-solar values in AGNs can give rise to the same feature even in lower accretion rate sources. Thus, we would expect two separate streams of material, at two different distances from the black holes.

\subsection{FRADO predictions vs. observations} \label{sec:observation}

As discussed above, our reasonably justified model is also fast enough compared to HD models to produce results for a large grid of initial physical parameters. We believe that the negligence of the hydro effects is not very important. As argued by \citet{Risaliti2010}, the motion in the radiative acceleration phase, and later the motion of the clump the clump in the medium are highly supersonic, so the effect of the pressure gradient to change the density and the internal pressure of the clouds can be neglected. The fact that our simple description recovers the motion complexity seems to support this view. Thus approximate description may be good enough for testing if indeed the radiation pressure acting on dust drives the motion of the BLR clouds.

\subsubsection{RL relation}

The FRADO model well predicts the basic location of the LIL BLR \citep{Czerny2011,Czerny2017}, and the estimates using the shielding allowed to recover the radius-luminosity (RL) relation including the dispersion \citep{Naddaf2020}. The earlier studies implied rather tight relation but new reverberation mapping results indicate considerable dispersion in RL relation \citep{CzernyTimedelay2019, Dupu2019, fonseca2020, Martinez2020}. This dispersion, apparently different depending on the broad-line adopted as the indicator \citep{zhang2021}, is most likely related to a spread in the Eddington ratio \citep[e.g.][]{Dupu2016, Naddaf2020, michal2020, michal2021}, which reflects the spread in the optical/UV SED and the available ionizing flux \citep{fonseca2020}. 

A preliminary test of our results with observation can be done as follows. For the high Eddington rate, we consider the most densely impacted radius with falling clouds, covering a broad range of aspect ratios i.e. $8000 R_{g}$ as shown in figure \ref{fig:single_global}(p and q) to be the location of BLR. As for the low Eddington rates, the location of BLR is taken to be where the highest peak is attained by clouds, most likely to be visible to the observer, which is around $2400 R_{g}$ and $1100 R_{g}$ as shown in figure \ref{fig:BLR_shape2} and \ref{fig:BLR_shape3} for rates of $0.1$ and $0.01$, respectively. We calculate the time-delays for the viewing angle of $i=39.2$ \citep{Lawrence2010} and for both the clouds located in the [closer - farther] side of AD relative to the observer using
\begin{equation}
\tau = \frac{R_{\mathrm{BLR}}}{c} \left( \sqrt{1+ q^2} \pm sin~i - q~ \cos~i \right)
\end{equation}
where $q$ is the aspect ratio as in the figure \ref{fig:single_global}(q). The value of $q$ is negligible for low Eddington rates, while for the high rate we take the median value of $q=0.15$. It gives $[4.57 - 20.42]$, $[10 - 44.67]$, and $[23.44 - 138]$ days for the Eddington rates of $0.01$, $0.1$, and $1$, respectively.
We can compare these limits with the transfer function for H$\beta$ line determined observationally. For example, the transfer function in Mrk 817, black hole mass of ($4.9 \pm 0.8) \times 10^7$ \citep{peterson2004}, and Eddington ratio 0.14 \citep{li2016} mostly peaks in the range of 10 - 30 days \citep{li2016}, which seems being consistent with our predictions given above for these accretion parameters. It must be stressed here again that in our model the location of the BLR depends only on the assumption of the dust sublimation temperature, and gives the location of the LIL part of BLR.
A detailed study on the RL relation and the transfer function measurements based on the model will be addressed later in the next paper.

\subsubsection{Line shapes and ratios}

In the current paper we do not yet address the issue of the line shapes and ratios since this requires a complex approach, mostly like in typical parametric models \citep[e.g.][]{pancoast2014}. Our model predicts the radiation flux seen by the clouds, but indeed the cloud density will have to be parametrized. We can start from assumption of constant cloud density since its quite universal value is supported by the line ratio fitting in LIL BLR \citep[e.g.][]{tek2016, panda2018, adhikari2019book}. Later we can limit the freedom by assuming the power law radial distribution of the density of the hot intercloud medium, and then determine the cloud density as a function of the radius from the pressure balance or pressure confinement \citep{rozanska2006,baskin2018}. We expect that our more complex dynamics will give a single-peak Lorentzian profiles not only for extremely low viewing angles \citep[e.g.][]{Goad2012} but also for moderate inclinations in the case of high Eddington ratio objects.

\subsubsection{BLR and torus}

It has been matter of debate due to many studies based on interferometry \citep[e.g.][]{clavel1989, Swain2003, Kishimoto2009, Pott2010, Kishimoto2011, Gravity2020} and based on RM \citep[e.g.][]{Suganuma2006, Koshida2014, Schnulle2015, Minezaki2019} whether dust is located within the BLR radii or it only sets the outer radius of BLR, defining the \emph{dusty torus}. Our new  sophisticated 3-D FRADO model predicts the launch of dusty material from the disk due to AD radiation pressure as the mechanism of formation of BLR whose trajectories can extend to large radii and high altitudes that can be responsible for the formation of torus. The results of complex flying material above the disk surface are consistent with the findings of studies by \citet{Goad2012, Figaredo2020}. It also shows that the emission of BLR and dusty region are interrelated as confirmed by \citet{Wang2013}.

The dusty torus has been introduced for masking the central disk from observation at high inclinations due to presence of a large amount of dust of high column densities extending from the equatorial plane to high altitudes \citep{antonucci1985}. The structure is not likely to be static, but a possible scenario involves the interaction between the outflows and inflows to form a geometrically thick torus \citep{wada2012}. \citet{Lawrence2010} suggested distorted/misaligned disks as the obscurer.

In most scenarios proposed so far for the formation of dusty torus, a sublimation radius is defined based on the condition of surviving dust from the spherical AD radiation field. The radius sets the onset of dusty torus, the first radius from which dust survives the intense radiation of whole AD at any altitude above the disk. However, the question is what brings the dusty material to those high altitudes. Therefore, a dynamical model of the wind was proposed for the torus itself \citep[e.g.][]{Konigl1994,Elitzur2006,gallagher2015}, but this was not a failed wind.
The dust location is constrained in an interesting observational study by \citet{markowitz2014} covering mostly high Eddington rate sources. It gives a radial domain spanning $0.3 - 140 \times 10^4 R_{g}$ for the location of BLR clouds. This implies that clouds are partially in the outer dusty torus, but partially in BLR. On the other hand, the outflow of BLR clouds towards the torus as in our results for high Eddington rate, and in other studies \citep[e.g.][]{Kawaguchi2010, Kawaguchi2011, Goad2012, honig2019, Figaredo2020} can provide some momentum to the claim that the dusty torus is a posterior to BLR and is indeed part of it; however this view is far from being widely and firmly accepted.

\subsubsection{The mass outflow rate}

In order to estimate the ejected mass from the AD due to radiation pressure within the \emph{escaping zone}, we assume that the clouds are optically thin at the time of launching. So the optical depth of the cloud, $\tau$, in the vertical direction can be of order of one. The Planck mean at sublimation radius is almost 50 times of the Thompson value (see figure \ref{fig:opacity}). One can obtain the column density of $N_{\mathrm{H}}^{\mathrm{vert}}= 3 \times 10^{22}$ [cm$^{-2}$] in the vertical direction, according to $\tau=N_{\mathrm{H}}^{\mathrm{vert}} \times \sigma_{\mathrm{Pl}}$ where $\sigma_{\mathrm{Pl}}$ is the Planck mean opacity. It is much higher than the column density in the horizontal direction discussed in Section~\ref{sect:3D}.  We are thus able to approximate the total flow of the vertically ejected mass as
\begin{equation}
\dot M_{\rm outflow} =
\frac{2 \pi\ R_{\mathrm{l}}\ \Delta R\ N_{\mathrm{H}}}{t_{\mathrm{exit}}} =
1.1152 \times 10^{23} ~~ \mathrm{[g/s]}
\end{equation}
where $R_{\mathrm{l}} = 5680 R_{g} $ is a launching radius within the \emph{escaping zone}, $\Delta R = 51 R_{g}$ the width of the \emph{escaping zone}, and $t_{\mathrm{exit}}= 1.79 \times 10^{8}$ [s] is the time it takes for the ejected material at $R_{\mathrm{l}}$ to leave the \emph{escaping zone}. Compared to the whole accretion rate of $\dot M_{\mathrm{edd}}= 1.399 \times 10^{26} ~~ \mathrm{[g/s]}$ for accretion rate of $\dot m = 1$ and the adopted $M_{\mathrm{BH}}=10^{8} M_{\odot}$, the value is small. Although it does not perturb the whole accretion process, it is not negligible either. However, this implies the modeled stream of material does not explain the BAL QSO flow \citep{Borguet2013} where the outflow is massive. So the huge amount of material must come from somewhere else, most probably some circumnuclear rings etc. 

\subsection{Number of clouds and probable collisions}\label{sec:collisions}

Taking an upper limit for mass loss rate of the entire disk at high Eddington rate, i.e. $10^{24}$  [g/s], one can simply estimate the total number of BLR clouds at any given moment, and also the mean flight time between two successive direct collision of clouds. The mass of BLR clouds of the typical size of $10^{12}$ to $10^{13}$ [cm] and typical density of $10^{12}$ up to $10^{13}$ [cm$^{-3}$] lies in the range of $10^{24}$ to $10^{28}$ [g] which corresponds to launching $10^{-4}$ up to $1$ cloud per second. Combining with flight time of clouds of around several tens of years (as implied by the figure \ref{fig:single_global}),
we obtain a total number of $10^{5}$ to $10^{9}$ clouds building BLR.

A simple formula from thermodynamics yields the collision time-scale of clouds as
\begin{equation}
\tau_{\rm coll} = \dfrac{~\lambda~}{\overline{v}}
= \dfrac{1}{\sqrt{2}~ n_{\rm cl}~ \sigma~ \overline{v}}
\end{equation}
where $\overline{v}$ is the average velocity of clouds which in this case is of order of $1000$ [km/s], and $\lambda$ is the {\it mean free path} where $n_{\rm cl}$ is the volume number density of clouds, and $\sigma$ is their collision cross-section.
It can be rewritten as
\begin{equation}
\tau_{\rm coll} \simeq ~0.1~
\dfrac{R^3_{\rm BLR}~n_{\rm H}~m_{\rm p}~R_{\rm cloud}}{\dot M_{\rm outflow}~ t_{\rm flight}~ \overline{v}}
\end{equation}
for a very geometrically flattened shallow BLR, with a height of order of 0.1 of its radial size. Recasting it for the BLR of the typical radial size of $10^{17}$ [cm] we have
\begin{equation}
    \tau_{\rm coll} \simeq 20
    \left(\frac{n_{\rm H}}{10^{12}\,{\rm cm^{-3}}} \right)
    \left(\frac{R_{\rm cloud}}{10^{12}\,{\rm cm}} \right)
    \left(\frac{100\,{\rm yr}}{t_{\rm flight}} \right) {\rm [yr]}
\end{equation}

This is almost the minimum value for the collision time-scale since the flight time of clouds are mostly less than 100 years (see figure \ref{fig:single_global}e), and also a minimum value for $n_{\rm H}$ or $R_{\rm cloud}$ and an upper limit for the disk mass loss rate are adopted. It therefore implies that the direct encounters are not highly probable so that the clouds may experience one collision during a full orbit. So neglecting the adjustment of a given cloud trajectory due to adjacent trajectories is a relatively safe assumption in our calculation.

\subsection{Dusty BLR studies in HD context}

Our description of the dynamics relying on the presence of dust and the important role of dust-driving in BLR, although simplified it shows good potentiality to address observational features as discussed in \ref{sec:observation}. However the results for the dynamics from our model cannot be compared yet with HD simulations since these simulations did not aim to address the dusty BLR. Instead, there is a long list of HD studies and/or advances physically-based studies aimed to model the dusty torus \citep[e.g.][]{Konigl1994, dorodnitsyn2012, wada2012, chan2016, chan2017, williamson2019, hoenig2019, Huang2020} or assuming the line-driven mechanism \citep[e.g.][]{murray1995, proga2000, higginbottom2014, waters2016, WatersProga2016, waters2021}. Torus-modelling papers assume the location of the inner radius of their structure at the dust sublimation temperature calculated from the total bolometric flux, so the temperature of the medium is lower than in the BLR. We know from observations that this region is by a factor of $\sim$ 5 larger than the BLR radius measured from the H$\beta$ line delay \citep{Koshida2014}. Papers based on line-driven winds predict the BLR radius too small for LIL like H$\beta$. For example, in \citet{proga2000} the outflowing stream starts at 7 light days (for a black hole mass $10^8$ $M_{\odot}$ and Eddington ratio 0.5), rather more appropriate for HIL lines \citep[also see ][]{WatersProga2016}, while our dust-based model gives 27 days, as expected for H$\beta$. We therefore have to look forward to future HD models incorporating dust-related mechanisms in BLR dynamics.

The future development of realistic HD models for the dusty BLR will be certainly difficult, as hinted by rather advanced dust/gas dynamics modelling done in the context protoplanetary disks \citep[see e.g.][]{vinkovi2021}. Another important issue in HD simulations is that the clumps in current simulations, either dusty or gaseous are of very low density, very large size and/or very far from the center \citep[e.g.][]{waters2021} compared to dense small BLR clouds of the size of $10^{12}$ up to $10^{13}$ cm \citep[e.g.][]{risaliti2011, pietrini2019} locating closer to the central irradiating source. Regarding the typical size of BLR of order of $10^{17}$ cm, one needs to have a spacial resolution of $\sim 10^{-4}$ or better in order to resolve a single BLR cloud, never yet reached by highly time- and computational-expensive 2D or 3D HD simulations. Alternatively,  we can just rely on the observational facts to test the model, and this is the path we plan to take in the nearest future.

\section{Summary}

We tested the dynamics of BLR under the 2.5-D non-HD prescription of the FRADO model.
In this test we incorporated the model with the wavelength-dependent dust opacities and two proposed configurations for the shielding effect.

As the results imply, the model is similar to the BL18 model of static puffed-up disk, although FRADO model catches the whole dynamical shape of BLR; most importantly, for high accretion rate where BL18 noticed the loss of static solution. The FRADO model also predicts the thin-funnel-like stream of escaping material proposed in EL00 model for high accretion rate.

The shape of BLR for high accretion rate seems to be very complicated which does not show any resemblance to usually adopted shapes for the BLR including disks, rings, shells, or cones. It is intuitively expected to produce single-peak emission profiles for high accretion rate. However, we expect to have disk-like lines profile (double-peak emission lines) for low accretion rates due to simple up/down motions of clouds with small vertical heights and velocities. We will calculate the shape of line profiles and examine their dependence on accretion rates and other parameters in the next paper.

The model shows that it may account for the LIL part of the BLR. We previously performed a preliminary test of the model with the radius-luminosity (RL) relation. It was successful in explaining the observed dispersion in H$\beta$ RL relation based on the Eddington ratio. In our next paper we will consider time-delay measurements resulting from the 3-D FRADO model in full details.

\acknowledgments
\section*{acknowledgments}

The project was partially supported by the Polish Funding Agency National Science Centre, project 2017/26/A/ST9/00756 (MAESTRO 9), and MNiSW grant DIR/WK/2018/12.
Authors would like to thank D. Semenov for helpful information he provided us on dust opacities. Authors are also grateful to the referee for the fruitful comments which considerably improved the quality of the paper.

\newpage

\appendix

\section{Physical calculations}\label{sect:appendix_A}

\subsection{Radiation pressure from an extended luminous disk acting on a single cloud}

The computations of the dusty cloud motion are based on the determination of the radiation pressure from the extended accretion disk. Our approach is basically similar to the computations done by \citet{icke1980}, but the medium opacity was there described by wavelength-independent Thomson cross-section for electron scattering, which reduces the problem considerably. We here use a complex description of the opacity, and the computation of the force from a given disk part depends not only on the relative geometrical position of the disk/cloud system, but also on the disk local temperature. Computations of the radiative force from the line driven wind in a number of papers \citep[e.g.][]{Pereyra1997,proga1998,feldmeier1999} were also done by integrating the force over the extended disk, but in this case no integration is performed over the wavelength for a specific contribution from the disk, and only the local disk flux is important, not the spectral shape. On the other hand, indeed line-driven force depends on the velocity gradient, which complicates considerably the dynamical computations. In the case of dust radiation pressure, no coupling with velocity is present, but the wavelength integration, grain size distribution and chemical composition is important as it allows automatically to include the UV and IR force component. We also take care of the effect of the scattering and absorption, separately. Since this integration over the wavelength is an important aspect of the computations, we present the details of the method below.

The time-independent radiation pressure due to absorption and scattering by definition \citep{mihalas1978book} are

\begin{figure}[b]
	\centering
	\includegraphics[scale=0.55]{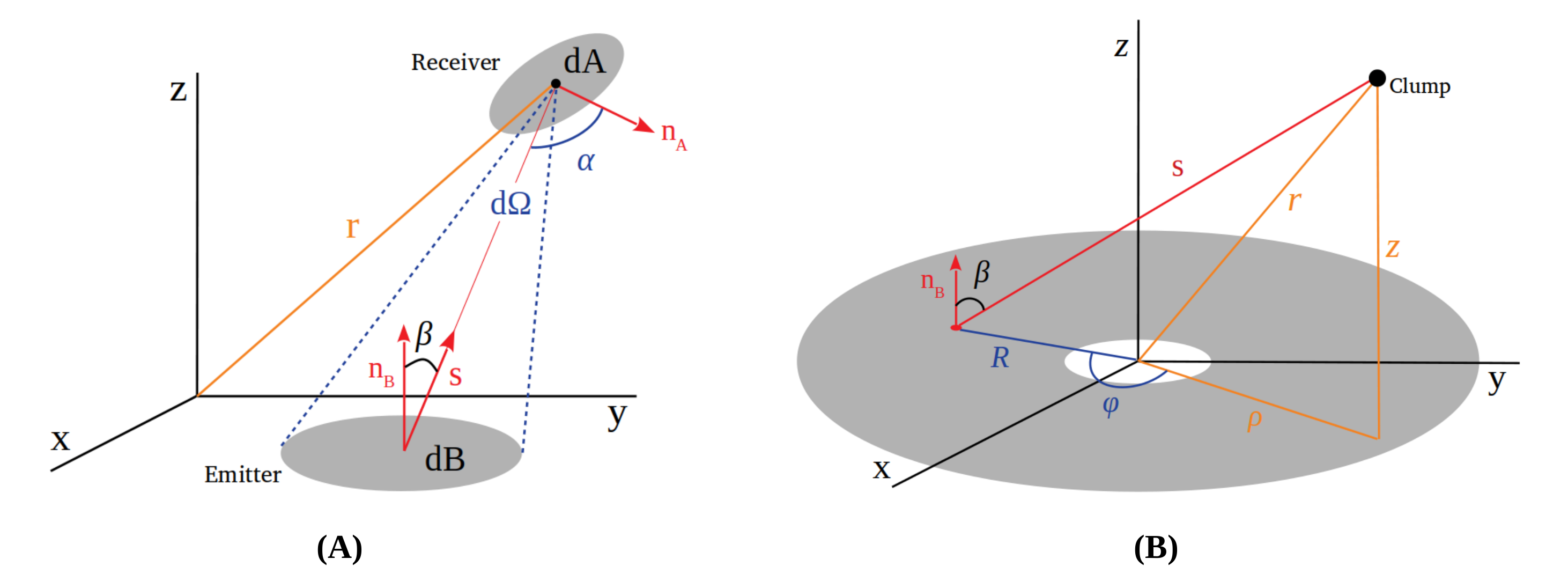}
	\caption{(A) Geometry of the emitter-receiver system which resembles the disk-cloud system for a differential (infinitesimal) segment of the disk. (B) Geometry of the disk-cloud system.}
	\label{fig:geometry_DiskCloud}
\end{figure}

\begin{equation}
P^{\mathrm{abs}} = \frac{1}{c} \int\int I_{\lambda}(\textbf{r}, \hat{\textbf{s}})\ \cos\alpha\ d\Omega\ d\lambda
\end{equation}

\begin{equation}
P^{\mathrm{sca}} = \frac{2}{c} \int\int I_{\lambda}(\textbf{r}, \hat{\textbf{s}})\ \cos^2\alpha\ d\Omega\ d\lambda
\end{equation}
where $I$ is the intensity of the radiating object, $c$ the speed of light, and $\alpha$ the angle between the unit vector $\hat{\textbf{s}}$ and the normal to the area element $\hat{\textbf{n}}_{A}$ shown in the figure \ref{fig:geometry_DiskCloud}(a). Therefore the absorption- and scattering-driven radiative acceleration of an infinitesimal object ($dA$) irradiated by an infinitesimal emitter ($dB$) shown in figure \ref{fig:geometry_DiskCloud}(a) are

\begin{equation}
\textbf{a}^{\mathrm{abs}} = \frac{1}{M c} \int\int I_{\lambda}\ \sigma_{\lambda}^{\mathrm{abs}}\ \cos\alpha\ \hat{\textbf{s}}\ d\Omega\ d\lambda
\end{equation}

\begin{equation}
\textbf{a}^{\mathrm{sca}} = \frac{2}{M c} (-\hat{\textbf{n}}_{A})
\int\int I_{\lambda}\ \sigma_{\lambda}^{\mathrm{sca}}\ \cos^2\alpha\ d\Omega\ d\lambda\ 
\end{equation}
where $M$ the mass of irradiated object (the receiver), and $\sigma_{\nu}$ is the cross section of the irradiated object for absorption or scattering which in general is frequency-dependent. It should be noted that we have put $\sigma_{\nu}$ rather than $dA$ which is applicable provided that the gradient of radiation pressure across the irradiated object (receiver) is negligible. It implies a small receiver (or, equivalently, grains with small values of cross-sections).

Assuming the effective cross section of the receiver to be always perpendicular to $\hat{\textbf{s}}$ so $\alpha=0$ and we have
\begin{equation}
\textbf{a}_{\mathrm{grain}}^{\mathrm{abs}} =
\frac{1}{M_{\mathrm{grain}}}  \frac{1}{c}
\int\int I_{\lambda}\ \sigma_{\lambda}^{\mathrm{abs}}\ \hat{\textbf{s}}\ d\Omega\ d\lambda
\end{equation}
\begin{equation}
\textbf{a}_{\mathrm{grain}}^{\mathrm{sca}} =
\frac{1}{M_{\mathrm{grain}}}  \frac{2}{c}
\int\int I_{\lambda}\ \sigma_{\lambda}^{\mathrm{sca}}\ \hat{\textbf{s}}\ d\Omega\ d\lambda
\end{equation}
so the total radiative acceleration of the object is
\begin{equation}
\textbf{a}_{\mathrm{grain}}^{\mathrm{rad}} =
\frac{1}{M_{\mathrm{grain}}}  \frac{1}{c}
\int\int I_{\lambda}\ \sigma_{\lambda}^{\mathrm{rad}}\ \hat{\textbf{s}}\ d\Omega\ d\lambda
\end{equation}
where
$
\sigma_{\lambda}^{\mathrm{rad}}=
\sigma_{\lambda}^{\mathrm{abs}} + 2 \sigma_{\lambda}^{\mathrm{sca}}
$.

Using the definition of solid angle, we consider an extended disk-like radiating surface shown in figure \ref{fig:geometry_DiskCloud}(a) in grey where red spot is the area element in polar coordinate, $dB$, shown in the figure \ref{fig:geometry_DiskCloud}(b) and $\hat{\textbf{n}}_{B}$ is the unit vector normal to $dB$. So we have
\begin{equation}
d\Omega = \frac{\cos\beta}{s^2}\  R\ dR\ d\varphi
\end{equation}
where $\cos\beta= z/s$, and $\hat{\textbf{s}}=\textbf{s}/s$.
So we have
\begin{equation}
\textbf{a}_{\mathrm{grain}}^{\mathrm{rad}} =
\frac{1}{M_{\mathrm{grain}}}  \frac{z}{c}
\int\int\int I_{\lambda}\ \sigma_{\lambda}^{\mathrm{rad}} \ \frac{{\textbf{s}}}{s^4}\  R\ dR\ d\varphi\ d\lambda
\end{equation}

Knowing that $\ (\mathbf{s=r-R})$, using the below definitions
\begin{equation}
\left\{
\begin{array}{lclcl}
\textbf{r}  = & \rho &\hat{\rho}\ + & z &\hat{\textbf{z}}   \\
\textbf{R}  = &  R\ \cos\varphi & \hat{\rho}\ + & R\ \sin\varphi &\hat{\rho}_{\bot}
\end{array}
\right.
\end{equation}
we can obtain the vector $\mathbf{s}$ and its length as
\begin{equation}
\textbf{s} =
(\rho- R\ \cos\varphi)\ \hat{{\rho}} -
R\ \sin\varphi\  \hat{{\rho}}_{\bot} +
z\ \hat{\textbf{z}}
\end{equation}
\begin{equation}
s^2= r^2 + R^2 - 2\ \rho\ R\ \cos\varphi
\end{equation}

Due to azimuthal symmetry, the summation over the second term of $\textbf{s}$ would give zero, so we have
\begin{equation}
\textbf{a}_{\mathrm{grain}}^{\mathrm{rad}} =
\frac{1}{M_{\mathrm{grain}}}  \frac{z}{c}
\int\int\int I_{\lambda}\ \sigma_{\lambda}^{\mathrm{rad}}\ 
\frac{(\rho- R\ \cos\varphi)\ \hat{{\rho}} + z\ \hat{\textbf{z}}}
{( r^2 + R^2 - 2\ \rho\ R\ \cos\varphi ) ^{2}}\  R\ dR\ d\varphi\ d\lambda
\end{equation}
for which one can have the components in Cartesian coordinates just using
$\hat{\rho}  = ( x\ \hat{\mathbf{x}} + y\ \hat{\mathbf{y}} ) / \rho$.

This is different from the approach of \citet{icke1980} in terms of the denominator. In the definition of $\hat{\textbf{s}}=\textbf{s}/s$, we can remove the second term of $\hat{\textbf{s}}$ due to azimuthal symmetry but it can not be removed from $s$ (the magnitude of $\hat{\textbf{s}}$).
However this mistake did not propagate in the literature.

This gives the net radiative acceleration for a single grain of a specific size and certain type of material. But for a dusty clump, i.e. a distribution of dust particles (with different material and different size) embedded in and strongly coupled with a volume of gas, the radiative acceleration can be obtained by summation over the type and size of dust particles as below

\begin{equation}
\label{eq:rad_pres1}
\textbf{a}_{\mathrm{clump}}^{\mathrm{rad}} =
\frac{1}{M_{\mathrm{clump}}}
\frac{z}{c}
\int\int\int
I_{\lambda}\  \sigma_{\lambda}^{\mathrm{tot(rad)}}\ 
\frac{(\rho- R\ \cos\varphi)\ \hat{{\rho}} + z\ \hat{\textbf{z}}}
{( r^2 + R^2 - 2\ \rho\ R\ \cos\varphi ) ^{2}}\  R\ dR\ d\varphi\ d\lambda
\end{equation}
where $M_{\mathrm{clump}}=M_{\mathrm{dust}}+M_{\mathrm{gas}}$.


The final form of radiative acceleration for the dusty clump is
\begin{equation}
\label{eq:rad_pres2}
\textbf{a}_{\mathrm{clump}}^{\mathrm{rad}} =
\frac{\Psi}{1+\Psi} \frac{z}{c}
\int_{\lambda_{i}}^{\lambda_{f}}
\int_{\varphi_{min}}^{\varphi_{max}}
\int_{R_{min}}^{R_{max}}
I_{\lambda}\  K^{\mathrm{rad}}_{\lambda}\
\frac{(\rho- R\ \cos\varphi)\ \hat{{\rho}} + z\ \hat{\textbf{z}}}
{\left( r^2 + R^2 - 2\ \rho\ R\ \cos\varphi \right) ^{2}}\  R\ dR\ d\varphi\ d\lambda
\end{equation}
where $\Psi$ is the clump dust-to-gas ratio, and $K^{\mathrm{rad}}_{\lambda}$ is the mean total cross-section per dust mass which resembles the definition of opacity (see section \ref{sec:append_dust} for details).

Incorporating with the radiation pressure on $w$-percent ionized gas, one can generally write
\begin{equation}
\textbf{a}_{\mathrm{clump}}^{\mathrm{rad}} =
\frac{1}{M_{\mathrm{clump}}} \frac{F}{c} \left( \sigma_{\mathrm{dust}}^{\mathrm{tot}} +
\sigma_{\mathrm{T}}^{\mathrm{tot}}
\right)
\end{equation}
which can be rewritten as
\begin{equation}
\textbf{a}_{\mathrm{clump}}^{\mathrm{rad}} =
\frac{\mathbb{F}}{c} \left( 
\frac{\Psi}{1+\Psi}\ K^{\mathrm{rad}}_{\mathrm{dust}}
+ \frac{w}{1+\Psi}\ \frac{\sigma_{\mathrm{T}}^{\mathrm{tot}}}{M_{\mathrm{ion.gas}}} \right) \approx
\frac{\mathbb{F}}{c} \left( 
\Psi\ K^{\mathrm{rad}}_{\mathrm{dust}}
+ w\ \frac{\sigma_{\mathrm{T}}}{m_{\mathrm{p}}} \right)
\end{equation}
so if the fist term dominates the second term, the radiation pressure due to Thompson electron scattering can be neglected (see figure \ref{fig:opacity}).

It should be noted that we have considered the strong-coupling approximation. So the clump is a unified rigid body within which the embedded spherical dust particles and the volume of gas are sort of wired or connected to each other. Also the clump is not of a large size or the dust particles are mostly concentrated around its center. Otherwise, randomly-oriented acceleration vectors of dust particles located at random positions within a large volume of gas will result in a different net acceleration vector. This becomes worst if there is no dust-gas strong-coupling which results in deformation, disintegration, and fragmentation of the clump.

\subsection{Dust opacity and dust-to-gas ratio}\label{sec:append_dust}

In order to find $\sigma_{\lambda}^{\mathrm{tot(rad)}}$, assuming a dust model with a population of dust types with a given grain size distribution, we proceed with a general relation valid for grains with radius $a_{-} \leq a \leq a_{+}$ as below
\begin{equation}\label{eq:distribution}
dn_{i}(a) = n f_{i}(a)\ da = n_{i}(a)\ da 
\end{equation}
where $n_{i}(a)$ is the number-density of grains and $n$ is the number-density of H nuclei ($n = n_{H} + 2 n_{H_{2}}$), $i$ stands for dust sort, and $f_{i}$ is a modular function.
Now one can write for example the total absorption cross-section (the same applies to total scattering cross-section) for a given dust sort as
\begin{equation}
\sigma_{\lambda, i}^{\mathrm{tot(abs)}} = V_{\mathrm{clump}}
\int^{a^{i}_{+}}_{a^{i}_{-}}
n_{i}(a)\ [\sigma(a)]_{\lambda, i}^{\mathrm{abs}}\ da
\end{equation}
where $[\sigma(a)]_{\lambda, i}^{\mathrm{abs}}$ is the absorption cross-section of a single grain of a certain sort of size $a$ (radius) at wavelength $\lambda$, and $V_{\mathrm{clump}}$ is the volume of the clump. Summation over sorts of dust gives
\begin{equation}
\sigma_{\lambda}^{\mathrm{tot(abs)}} =
\sum_{i=1}^{\mathrm{NDS}}
\sigma_{\lambda, i}^{\mathrm{tot(abs)}} =
V_{\mathrm{clump}}
\sum_{i=1}^{\mathrm{NDS}} \int^{a^{i}_{+}}_{a^{i}_{-}}
n_{i}(a)\ [\sigma(a)]_{\lambda, i}^{\mathrm{abs}}\ da
\end{equation}
where $\mathrm{NDS}$ is the number of dust sorts in the dust model. Likewise we can find $\sigma_{\lambda}^{\mathrm{tot(sca)}}$ and consequently $\sigma_{\lambda}^{\mathrm{tot(rad)}}$.

When calculating the radiative acceleration, it would be easier to work with general densities, ratios, and mean values rather than certain masses or volumes. Looking at the equation \ref{eq:rad_pres1} we can write
\begin{equation}
\frac{\sigma_{\lambda}^{\mathrm{tot(rad)}}}{M_{\mathrm{clump}}} =
\frac{M_{\mathrm{dust}}}{M_{\mathrm{clump}}}\
\frac{\sigma_{\lambda}^{\mathrm{tot(rad)}}}{M_{\mathrm{dust}}} =
\frac{\Psi}{1+\Psi}\ K^{\mathrm{rad}}_{\lambda}
\end{equation}
where $K^{\mathrm{rad}}_{\lambda}$ is the total opacity of the clump given by
\begin{equation}\label{eq:K_value1}
K^{\mathrm{rad}}_{\lambda}= K^{\mathrm{abs}}_{\lambda} + 2 K^{\mathrm{sca}}_{\lambda}
\end{equation}
where
\begin{equation}\label{eq:K_value2}
K^{\mathrm{j}}_{\lambda}= \frac{1}{\Psi}
\sum_{i=1}^{\mathrm{NDS}} K^{\mathrm{j}}_{\lambda,i} \Psi_{i}
\end{equation}
where
\begin{equation}\label{eq:K_value3}
K^{\mathrm{j}}_{\lambda, i} =
\frac{\sigma_{\lambda, i}^{\mathrm{tot(j)}}} {M_{\mathrm{dust}, i}} =
\frac{1}{\rho_{\mathrm{dust}, i}}\ 
\int^{a^{i}_{+}}_{a^{i}_{-}}
n_{i}(a)\ [\sigma(a)]_{\lambda, i}^{\mathrm{j}}\ da
\end{equation}
where $j$ stands for ($\mathrm{rad}$), ($\mathrm{abs}$), or ($\mathrm{sca}$); and $\Psi$ is dust-to-gas mass ratio given by
\begin{equation}\label{eq:dust-to-gas}
\Psi = \frac{M_{\mathrm{dust}}}{M_{\mathrm{gas}}} =
\frac{\rho_{\mathrm{dust}}}{\rho_{\mathrm{gas}}} =
\sum_{i=1}^{\mathrm{NDS}} \Psi_{i} =
\frac{\sum_{i=1}^{\mathrm{NDS}} \int^{a^{i}_{+}}_{a^{i}_{-}}
n_{i}(a)\ 4/3 \pi a^3 \rho_{\mathrm{b}, i}\ da}{n m_{H}} =
\frac{\sum_{i=1}^{\mathrm{NDS}} \int^{a^{i}_{+}}_{a^{i}_{-}}
f_{i}(a)\ 4/3 \pi a^3 \rho_{\mathrm{b}, i}\ da}{m_{H}}
\end{equation}
where $\rho_{\mathrm{b}, i}$ is the bulk dust density of sort $i$, and $m_{H}$ is the mass of H nuclei.

\subsection{Dust sublimation}\label{sec:criterion}

Assuming instantaneous re-emission of the absorbed radiation by dust in the form of an isotropic blackbody radiation, if the amount of heat absorbed by dusty content of the clump increases its temperature to that of sublimation, the radiative engine of motion of the clump switches off. So the criterion for sublimation of dusty content is
\begin{equation}\label{eq:criterion}
E_{\mathrm{abs}} = E_{\mathrm{emit}}(T_{\mathrm{s}})
\end{equation}

The total radiative energy by definition \citep{mihalas1978book} is
\begin{equation}
E = \int I_{\lambda}(\textbf{r}, \hat{\textbf{n}})\ \cos(\alpha)\ d\lambda\ dA\ d\Omega\ dt
\end{equation}

So the radiative energy absorbed by dust is
\begin{equation}
E_{\mathrm{abs}} = \int I_{\lambda}\ \sigma^{\mathrm{tot(abs)}}_{\lambda}\ d\lambda\ d\Omega\ dt
\end{equation}
and the energy re-emitted by dust \citep{Loska1993} at the sublimation temperature ($T_{\mathrm{s}}$) is
\begin{equation}
E_{\mathrm{emit}}(T_{\mathrm{s}}) = 4 \pi \int B_{\lambda}(T_{\mathrm{s}})\ \sigma^{\mathrm{tot(abs)}}_{\lambda}\ d\lambda\ dt
\end{equation}

Dividing both sides of the equation \ref{eq:criterion} by $dt$ and $M_{\mathrm{dust}}$ we can introduce $Q$ (the total power per dust mass) so we have
$Q_{\mathrm{abs}} = Q_{\mathrm{emit}} (T_{\mathrm{s}})
$ to be the sublimation criterion where
\begin{equation}
Q_{\mathrm{emit}} (T_{\mathrm{s}}) = 4 \pi \int B_{\lambda}(T_{\mathrm{s}})\ K^{\mathrm{abs}}_{\lambda}\ d\lambda
\end{equation}
\begin{equation}
Q_{\mathrm{abs}} = \int I_{\lambda}\ K^{\mathrm{abs}}_{\lambda}\ d\lambda\ d\Omega
\end{equation}
so we have
\begin{equation}
\label{eq:Qabs}
Q_{\mathrm{abs}} = z
\int_{\lambda_{i}}^{\lambda_{f}}
\int_{\varphi_{min}}^{\varphi_{max}}
\int_{R_{min}}^{R_{max}}
\frac{I_{\lambda}\  K^{\mathrm{abs}}_{\lambda}}
{\left( r^2 + R^2 - 2\ \rho\ R\ \cos\varphi \right) ^{3/2}}\  R\ dR\ d\varphi\ d\lambda
\end{equation}

\bibliographystyle{aasjournal}
\bibliography{naddaf}

\end{document}